\documentclass[12pt]{article}
\usepackage{latexsym}
\usepackage{amsthm}
\usepackage{amsmath}
\usepackage{amsfonts}
\usepackage[dvips]{graphicx}
\usepackage{amssymb}
\usepackage{axodraw}
\usepackage{epstopdf}
\usepackage{epsfig}
\usepackage{pstricks}
\usepackage{color}

\newcommand{\vev}[1]{\ensuremath{\langle #1 \rangle}}
\newcommand{\LL}{\mathcal{L}}

\newcommand{\OO}{\mathcal{O}}
\newcommand{\epm}{e^+e^-}

\newcommand{\be}{\begin{eqnarray}}
\newcommand{\bea}{\begin{eqnarray}}
\newcommand{\ee}{\end{eqnarray}}
\newcommand{\eea}{\end{eqnarray}}

\newcommand{\f}{\frac}

\newcommand{\mWm}{m_{W_{D,\rm{min}}}}
\newcommand{\mhm}{m_{h_{D,\rm{min}}}}
\begin{document}
\begin{titlepage}
\begin{flushright}
SLAC-PUB-13561\qquad \\
SU-ITP-09/12\qquad\ \\
\end{flushright}
\vspace{.0in}

\begin{center}
\vspace{1cm}

{\Large \bf Probing Dark Forces and Light Hidden Sectors at Low-Energy $e^+e^-$ Colliders}

\vspace{0.8cm}

{\bf Rouven Essig$^1$, Philip Schuster$^1$, Natalia Toro$^2$}

\vspace{0.5cm}
{\it $^1$ Theory Group, SLAC National Accelerator Laboratory, \\
Menlo Park, CA 94025, USA}

\vspace{0.2cm}
{\it $^2$ Stanford Institute for Theoretical Physics, Stanford University, \\
Stanford, CA 94305, USA}

\end{center}
\vspace{1cm}

\begin{abstract}
A dark sector --- a new non-Abelian gauge group Higgsed or confined
near the GeV scale --- can be spectacularly probed in low-energy
$e^+e^-$ collisions.  A low-mass dark sector can explain the annual
modulation signal reported by DAMA/LIBRA and the PAMELA, ATIC, and
INTEGRAL observations by generating small mass splittings and new
interactions for weak-scale dark matter.  Some of these observations
may be the first signs of a low-mass dark sector that collider
searches can definitively confirm.  Production and decay of
$\OO$(GeV)-mass dark states is mediated by a Higgsed Abelian gauge
boson that mixes kinetically with hypercharge.  Existing data from
BaBar, BELLE, CLEO-c, and KLOE may contain thousands of striking
dark-sector events with a high multiplicity of leptons that
reconstruct mass resonances and possibly displaced vertices.  We
discuss the production and decay phenomenology of Higgsed and confined
dark sectors and propose $e^+e^-$ collider search strategies.  We also
use the DAMA/LIBRA signal to estimate the production cross-sections
and decay lifetimes for dark-sector states.

\end{abstract}

\end{titlepage}

\tableofcontents
\section{Introduction}
Low-mass particles neutral under the Standard Model are nearly
unconstrained by existing searches.  We consider low-energy collider
signatures of a new $\OO$(0.1-10 GeV)-mass ``dark'' sector with gauge
group $G_D$ and light matter charged under it.  The discovery of a
dark sector that extends the structure of known low-energy gauge
interactions would open a new frontier for particle physics and
provide a second laboratory for fundamental questions including gauge
unification and supersymmetry breaking.

\begin{figure}[t]
\begin{center}
\includegraphics[width=5in]{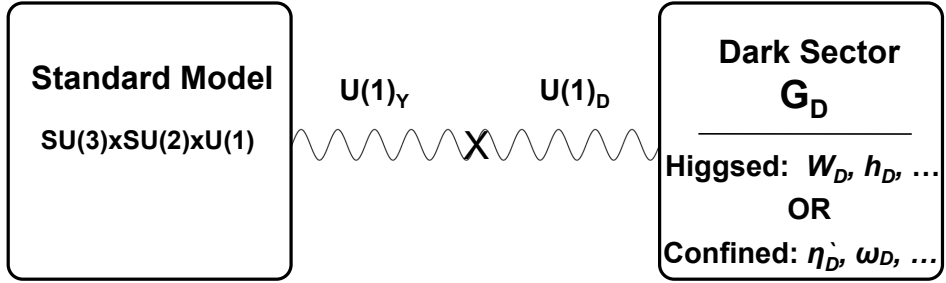}
\end{center}
\caption{
\label{fig:twoSectors} We consider a dark
sector with non-Abelian gauge group $G_D$, which is Higgsed or
confined at $\mathcal{O}(\mbox{MeV}-10~\mbox{GeV})$.  We assume that
$G_D$ contains a Higgsed Abelian factor $U(1)_D$, so that the dark
sector interacts with Standard Model matter through kinetic mixing of
hypercharge with the $U(1)_D$ gauge boson $A'$, of mass in the same
range.  Either the Higgsed or confined phases of $G_D$ necessarily include
new states that can be produced through $A'$ interactions.\label{fig:darkSector}}
\end{figure}

If $G_D$ contains an Abelian subgroup $U(1)_D$, high-scale physics
generically induces kinetic mixing between the dark $U(1)_D$ and
hypercharge, as sketched in Figure \ref{fig:twoSectors}; this mixing
mediates direct production of dark-sector matter in high-luminosity
scattering experiments.  The decay of the dark-sector matter can lead
to spectacular events with high lepton and hadron multiplicities,
lepton pair masses reconstructing resonances in the dark sector, and
possibly displaced vertices.  In optimistic scenarios, thousands of
these events could be sitting undiscovered in data collected by BaBar,
BELLE, CLEO-c and KLOE.  Reconstructing new resonances in these events
would reveal the dynamics of the dark sector.

Evidence for a low-mass dark sector is emerging from a surprising
source: accumulating hints from terrestrial and satellite dark matter
experiments indicate that dark matter is not an afterthought of the
Standard Model's hierarchy problem, but instead has rich dynamics of
its own.  The local electron/positron excesses reported by HEAT
\cite{Barwick:1997ig}, PAMELA \cite{Adriani:2008zr,Adriani:2008zq},
PPB-BETS \cite{Torii:2008xu}, ATIC \cite{2008zzr}, and others
\cite{Aguilar:2007yf,Grimani:2002yz} are suggestive of weak-scale dark
matter interacting with a new light boson, for which the $U(1)_D$ is a
natural candidate if its mass is $\OO$(GeV).  Dark-sector interactions
can also generate the mass splittings among dark-matter states
suggested by other experiments
\cite{ArkaniHamed:2008qn,Schuster:2009}.  Dark matter scattering
inelastically into an excited state split by $\OO$(100 keV) can
simultaneously explain the annual modulation signal reported by
DAMA/NaI \cite{Bernabei:2005hj} and DAMA/LIBRA \cite{Bernabei:2008yi}
and the null results of other direct-detection experiments
\cite{TuckerSmith:2001hy,Chang:2008gd}.  Likewise, the INTEGRAL 511
keV excess at the galactic center appears consistent with the
excitation of dark matter states, but requires a slightly larger
splitting of $\OO$(MeV) \cite{Strong:2005zx,Finkbeiner:2007kk}.

The electron/positron excesses and the splittings suggested by the
DAMA and INTEGRAL signals independently motivate an $\OO$(GeV)-mass
dark sector.  If \emph{any} of these anomalies are signals of dark
matter, the new dynamics required to explain them can be discovered at
$\epm$ colliders.  Among the anomalies, DAMA's signal offers the
most precise predictions for $\epm$ collider physics: the scattering
rate is sensitive to the strength of kinetic mixing between the
Standard Model and the dark sector, and gives reason to expect an
observable direct production cross-section for the dark sector.

\subsection*{Outline}
In the remainder of this introduction, we further develop the
motivation for a kinetically mixed light dark sector, and briefly
describe the resulting events in low-energy $e^+e^-$ collisions. 
In Section \ref{sec:mixingIntro}, we discuss the kinetic mixing that
couples the Standard Model to the dark sector.  In Section
\ref{sec:dmIntro}, we summarize the evidence that dark matter, with a
mass of 100 GeV to 1 TeV, may interact with a low-mass dark sector.
In particular, we introduce two frameworks for inelastic dark matter
that motivate the Higgsed and confined dark sectors considered
here. Though the dark matter in each model is heavy, in either case
the generation of inelastic splittings requires \emph{light} matter in
the dark sector that can be produced at $e^+e^-$ colliders.  We
describe the general structure of dark-sector production events in
Section \ref{sec:eventsIntro}.

We then work our way into the dark sector, and back out.  Two
production modes are of interest, and their cross-sections are nearly
independent of the detailed structure of the dark sector.  We present
the production cross-sections for general parameters in Section
\ref{subsec:production}, and for the parameter ranges motivated by the
inelastic dark matter models in Section \ref{sec:DAMANormalization}.  In
Section \ref{sec:generalSector}, we discuss the decay modes of
metastable states, and the shapes of events expected within either
Higgsed or confined dark sectors.  

Section \ref{sec:searches} contains a brief discussion of search
regions and strategies, which synthesizes the experimental
consequences of the results of Section \ref{sec:generalSector}.
Section \ref{sec:searches} is self-contained, and readers who are not
interested in the detailed decay phenomenology of different
dark-sector models can skip Section \ref{sec:generalSector}, and only
read this section.  We conclude in Section \ref{sec:conclusion}.

Appendix
\ref{app:Contraints} reviews the existing constraints on a new
$U(1)_D$ gauge boson with mass $\OO$(GeV) kinetically mixing with
hypercharge.  All numerical results are presented both for arbitrarily
chosen fixed couplings, and for couplings normalized to the DAMA/LIBRA
inelastic dark matter scattering cross-section using the prescription in Section
\ref{sec:DAMANormalization} and Appendix \ref{app:iDMreview}.

The B-factory phenomenology of pure $U(1)$ dark sectors has been
discussed in \cite{Batell:2009yf}, which notes a multi-lepton final
state from a ``Higgs$'$-strahlung'' process, and the phenomenology of
MeV-scale dark-matter models in which $\gamma+e^+e^-$ or
$\gamma+\mbox{nothing}$ final states dominate has been considered in
\cite{Borodatchenkova:2005ct}.  The discussion of Section 3 is similar
in spirit to the hadron collider phenomenology for Hidden Valleys
\cite{Strassler:2006im,Strassler:2006qa,Han:2007ae,Strassler:2008bv} and 
the Higgsed dark-sector models discussed in 
\cite{ArkaniHamed:2008qn,ArkaniHamed:2008qp,Baumgart:2009tn} (see also \cite{Hambye:2008bq}).
Other work that discusses new physics beyond the Standard Model leading to 
rare $\Upsilon$ decays at low-energy $e^+e^-$ colliders 
(albeit with very low-multiplicity or even 
no leptons in the final state), such as $\Upsilon\to\gamma+\mbox{nothing}$, 
$\Upsilon\to\mbox{invisible}$,  
or $\Upsilon\to \gamma H$ (where $H$ is a light Higgs boson),
includes \cite{Gunion:2005rw,McElrath:2005bp,Dermisek:2006py,Galloway:2008yh}. 

\subsection{$U(1)$ Mixing with Dark Sectors}\label{sec:mixingIntro}
Any new Abelian gauge group $U(1)_D$ has a gauge-invariant kinetic
mixing term with Standard Model hypercharge  \cite{Holdom:1986eq,Dienes:1996zr},
\be
\Delta \mathcal{L} = \epsilon_Y F^{Y,\mu\nu}F^{D}_{\mu\nu}. \label{eq:mixing}
\ee
Several high-scale mechanisms, summarized below, generate $\epsilon_Y \sim
10^{-6}-10^{-2}$ (we note that from string theory the 
range for $\epsilon_Y$ can be even larger \cite{Abel:2008ai}).  
Therefore, the setup illustrated in  Figure
\ref{fig:darkSector} is generic if there is any new gauge group 
containing a $U(1)_D$ factor.  

The operator \eqref{eq:mixing} is not
generated at the Planck scale if $U(1)_Y$ is embedded in a Grand
Unified Theory (GUT), because it is not $SU(5)$ gauge-invariant.
However, Planck-suppressed gauge-invariant operators involving
GUT-breaking Higgses generate mixing
\be
\epsilon_{tree} \sim \left(\f{M_G}{M_{Pl}}\right)^p
\ee
when these Higgses obtain vacuum expectation values, where the power
$p$ depends on the representation content of the Higgses.  For
example, $p=1$ for the operator $\f{1}{M_{\rm Pl}}\mathrm{Tr}[\Phi
  F^5_{\mu\nu}]F_D^{\mu\nu}$, where $\Phi$ is an $SU(5)$ adjoint
Higgs, and $F^5$ and $F_D$ are the $SU(5)$ and dark sector field
strengths, respectively.  

Kinetic mixing is also generated by heavy split multiplets 
charged under both $U(1)_D$ and $SU(5)$.  Integrating out these
multiplets generates loop-level mixing 
proportional to the logarithm of mass ratios within the
multiplet:
\be
\epsilon_{loop} \sim \f{g_1 g_D}{16\pi^2} N_f \log \left(\f{M}{M'}\right).
\ee
The mass splittings can themselves be generated by either tree- or loop-level
effects, leading to $\epsilon_Y$ suppressed by one or two loop factors.  

At low energies, the kinetic mixing can be removed by a field
redefinition of $A^{Y,\mu}$, inducing $\epsilon$-suppressed
couplings of the $A'$ to electromagnetically charged states,
\be
\LL \supset \epsilon g_D A'_\mu J^\mu_{EM},
\ee
where $\epsilon\equiv \epsilon_Y \cos\theta_W$ and $\theta_W$ is the
Weinberg weak mixing angle.  Therefore, a light $U(1)_D$ gauge boson can and
naturally does open a path into and back out of any dark sector.

\subsection{Dark Matter Motivation for a {\it Light} Sector}\label{sec:dmIntro}

Searches for a GeV-mass dark sector are particularly motivated by
recent evidence for electroweak-mass dark matter with non-trivial
structure.  The consistency of \emph{several} independent observations
of distinct phenomena with the same qualitative framework is striking.
Moreover, if any one of these observations is found to have a
non-dark-matter origin, a low-mass dark sector is still strongly
motivated by the other data.

Satellite observations and terrestrial experiments provide two
kinds of evidence for a low-mass dark sector: 
\begin{enumerate}
\item Positron and/or electron flux measurements by PAMELA \cite{Adriani:2008zr,Adriani:2008zq}, PPB-BETS \cite{Torii:2008xu} and
  ATIC \cite{2008zzr} point to an unknown local source of
  high-energy (100 GeV-TeV) electrons and positrons.  If their source is dark matter
  annihilation or decay, synchrotron radiation from these electrons and positrons
  could also explain the WMAP haze near the galactic center \cite{Finkbeiner:2003im}.  Two
  features of these experiments are suggestive of light states coupled
  to dark matter:
\begin{itemize}
\item The excess of electrons and positrons, without a visible
  anti-proton excess \cite{Adriani:2008zq}, suggests dominantly leptonic decay or
  annihilation channels.  One way to assure this is if the dark matter
  decays or annihilates into $\OO$(GeV) particles that are kinematically
  forbidden from decaying to baryons \cite{ArkaniHamed:2008qn}.
\item If the excesses are produced by dark matter annihilation, the
  annihilation cross-section required to explain the signal is
  10-1000 times larger than the thermal freeze-out cross-section. This can
  be explained by Sommerfeld enhancement of the annihilation
  cross-section through an $\OO$(GeV) mass force mediator at low velocities \cite{ArkaniHamed:2008qn}.
\end{itemize}
\item As we will discuss below, non-Abelian gauge interactions Higgsed or confined at the GeV scale
can generate small mass splittings among states charged under them.  Two observations
suggest that dark matter may have such splittings:
\begin{itemize}
\item The INTEGRAL telescope \cite{Strong:2005zx} has reported a 511
  keV photon signal near the galactic center, indicating a new source
  of $\sim$ 1-10 MeV electrons and positrons.  This excess could be
  explained by annihilation of light $\OO$(1-10 MeV) dark matter
  \cite{Boehm:2003bt}, or by $\OO$(100 GeV-1 TeV) dark matter with
  $\OO$(MeV) excited states \cite{Finkbeiner:2007kk}.  In the latter
  case, dark matter excited by scattering decays back to the ground
  state by emitting a soft $e^+e^-$ pair.
\item The DAMA/NaI \cite{Bernabei:2005hj} and DAMA/LIBRA
  \cite{Bernabei:2008yi} experiments have reported an annual
  modulation signal over nearly eleven years of operation.  Modulation
  is expected because the Earth's velocity with respect to the dark
  matter halo varies as the Earth moves around the sun, and the phase
  of the observed modulation is consistent with this origin.  A simple
  hypothesis that explains the spectrum and magnitude of the signal,
  and reconciles it with the null results of other experiments, is
  that dark matter-nucleus scattering is dominated by an inelastic
  process,
\be \chi\;N \rightarrow \chi^* \;N, \ee in which the dark matter
$\chi$ scatters off a nucleus $N$ into an excited state $\chi^*$ with
mass splitting $\delta \approx 100$ keV \cite{TuckerSmith:2001hy}.
\end{itemize}
\end{enumerate}

Among these hints of new dark matter interactions, the inelastic dark
matter (iDM) interpretation \cite{TuckerSmith:2001hy} of the
DAMA/LIBRA signal deserves particular attention, because it is
sensitive to the strength of coupling between the dark sector and the
Standard Model.  While the other data discussed above is consistent
with extremely small kinetic mixing $\epsilon$, the rate of scattering
at DAMA/LIBRA fixes $\epsilon$ within a factor of few for fixed model
parameters and $A'$ mass, in a typical range $10^{-7}\lesssim \epsilon
\lesssim 10^{-2}$.  Therefore, the spectrum and rate of the signal
reported by DAMA/LIBRA can be translated into predictions for the
couplings of light states observable at $e^+e^-$ colliders, as we will
discuss below and in Section \ref{sec:DAMANormalization}.

We summarize here how iDM reconciles the DAMA/LIBRA signal with null
results from other searches.  Inelastic scattering can only occur when
the center-of-mass kinetic energy of the dark-matter-nucleus system
exceeds the splitting $\delta$. This implies that for a given target
nucleus mass $m_N$ and recoil energy $E_R$, the minimum lab-frame
velocity of dark matter required for inelastic scattering is \be
\beta_{\rm min}(E_R) \simeq \f{1}{\sqrt{2 m_N E_R}} \left( \f{m_N
  E_R}{\mu} + \delta \right),
\label{eq:betaMinInel}
\ee
where $\mu$ is the dark-matter-nucleus reduced mass.
This energy threshold enhances the sensitivity of the DAMA/LIBRA experiment
over other direct detection experiments in three ways:
\begin{enumerate}
\item
 Dark matter with velocity below $v_{\rm min} \simeq \sqrt{2\delta/\mu}$ does
 not scatter. This minimum velocity increases for small $m_N$, thereby enhancing scattering off
 heavy nuclei such as Iodine used in DAMA/LIBRA relative to scattering off Silicon
 and Germanium at CDMS \cite{Ahmed:2008eu}.
\item Dark matter direct searches obtain the tightest bounds on elastically scattering
dark matter from low-energy nuclear recoils.  If dark matter scatters inelastically,
low-energy recoils require \emph{higher} dark matter velocity, and are suppressed
--- see equation (\ref{eq:betaMinInel}).
\item The fraction of the halo velocity profile that exceeds $v_{\rm
  min} \simeq \sqrt{2\delta/\mu}$ is exponentially sensitive to
  changes in the Earth's velocity, leading to an $\OO$(1)
  annual modulation of the scattering rate, in contrast to the $\simeq
  2-3\%$ modulation typical of elastic scattering, which arises from
  the simple linear dependence of the dark matter flux on the Earth's
  velocity.
\end{enumerate}

In this paper, we will consider dark-sector gauge groups Higgsed or confined
at the GeV scale, both of which can generate $\OO$(100 keV) splittings
among $\OO$(100 GeV-TeV) dark matter states, with a kinetically mixed $U(1)_D$
mediating inelastic scattering off nuclei \cite{ArkaniHamed:2008qn,Schuster:2009}.
If the dark gauge group $G_D$ contains a Higgsed non-Abelian factor, radiative
effects can split all components of matter charged under $G_D$, with splittings
\be
\delta \sim \alpha_D \Delta m_{W_D} \qquad \mbox{(Higgsed/radiative)},
\ee
where $\Delta m_{W_D}$ is a splitting of dark-sector gauge boson
masses \cite{ArkaniHamed:2008qn}.  For example, $\Delta m_{W_D} \sim
m_{W_D} \sim 1$ GeV and $\alpha_D\sim 10^{-4}$ gives $\delta\sim 100$
keV.

If instead a non-Abelian factor of $G_D$ confines at a scale
$\Lambda_D$, a heavy-flavor meson can be cosmologically long-lived and
thus a dark matter candidate \cite{Schuster:2009}.  A particularly
interesting limit for inelastic dark matter is when one dark quark is
heavy ($m_{Q_D} \approx m_{DM} \approx (100 \mbox{ GeV}-1 \mbox{ TeV}) \gg \Lambda_D$), and the other light ($m_{q_D} \lesssim
\Lambda_D$).  A cosmologically long-lived $Q_D$ forms uncolored hadrons
after confinement, of which the lightest, a spin-0 $Q_D \bar q_D$
meson, is a viable dark matter candidate.  The spin-1 excited $Q_D
\bar q_D$ meson is split by hyperfine interactions, which give
\be
\delta \sim \f{\Lambda_D^2}{N_C m_{DM}} \qquad \mbox{(confined/hyperfine)},
\ee
where $N_C$ is the number of colors of $G_D$.  For example,
$m_{DM}\sim 1$ TeV, $N_C=3$, and $\Lambda_D\sim 500$ MeV gives $\delta \sim 100$
keV.

 For both Higgsed and confined scenarios, the $U(1)_D$ described above
 in Section \ref{sec:mixingIntro} can mediate inelastic scattering of
 dark matter off Standard Model nuclei, with elastic scattering
 naturally suppressed.  These mechanisms can also generate a splitting
 of $\OO$(MeV) to explain the INTEGRAL excess \cite{Finkbeiner:2007kk}.

Remarkably, though the mechanisms for generating the splitting from Higgsed and confined dark
sectors are quite different, both are suggestive of dark sectors with multiple new states at $\OO$(GeV)
--- precisely the range accessible in a variety of low energy $e^+e^-$ machines.
In the Higgsed case, these states include gauge and Higgs bosons.
In the confined case, these states include glueballs, light-flavor
mesons, and baryons.  Therefore, if either scenario
is correct, experiments such as BaBar, BELLE, CLEO-c, and KLOE
may be capable of discovering the kinds of spectacular events
that we consider here.

We will see that the combination $\alpha_D\epsilon^2/m_{A'}^4$, where
$\alpha_D=g_D^2/4\pi$, determines the DAMA/ LIBRA modulation rate.
Therefore, we can use the observed modulation signal to estimate
$\alpha_D\epsilon^2$ as a function of $m_{A'}$, which in turn
determines the expected production cross-section for $U(1)_D$-charged
matter at $e^+e^-$ colliders.  We discuss this estimate in Section
\ref{sec:DAMANormalization}.  The expected cross-sections are
$\OO$(fb) for $m_{A'} \sim $ GeV.  For heavier $A'$, larger mixing
parameters are required to reproduce the DAMA/LIBRA scattering rate; 
the resulting expected production cross-sections scale as $\sim m_{A'}^4$.
Therefore, very large cross-sections are expected for a few-GeV $A'$,
while a dark sector with sub-GeV $A'$ mass may evade detection in
existing data.

\subsection{Striking Events at B-factories}\label{sec:eventsIntro}
Existing B-factory datasets, containing over 1.4 ab$^{-1}$ of data
altogether, are ideally suited to search for a dark $U(1)_D$ in
the $\sim$ 100 MeV -- 10 GeV mass range, and for any dark sectors they connect
to. There are two leading production processes: an $A'$ with mass
beneath 10 GeV can be produced on-shell in association with a photon
(the radiative return process).
The $A'$ initiates a cascade of dark sector decays
if kinematically allowed; otherwise it decays to Standard Model
fermions or hadrons. In addition, any light charged states within
the dark sector can be produced through an off-shell $A'$ carrying
the full center-of-mass energy of the $e^+e^-$ collision.
The signatures of these processes are quite spectacular because decays
or parton showers within the dark sector generate high particle multiplicities.

\begin{figure}[t]
\begin{center}
\includegraphics[width=5in]{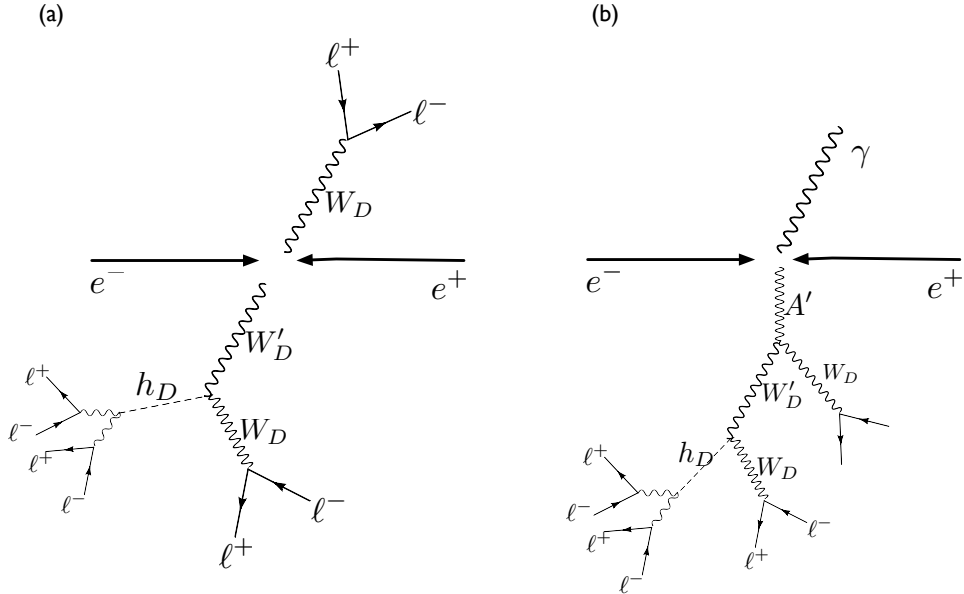}
\end{center}
\caption{Left: Cartoon of an event in which non-Abelian gauge bosons
  in a Higgsed dark sector are produced through an off-shell $A'$: the
  gauge bosons may decay into other dark bosons (or fermions if
  present), which in turn decay to light Standard Model fermions.
  Right: a similar final state recoils off a hard photon in
  $A'+\gamma$ radiative return production. \label{fig:higgsedCartoon}}
\end{figure}

We focus on two cases, motivated by the dark matter models discussed
in Section \ref{sec:dmIntro}: a Higgsed dark sector containing only the
non-Abelian gauge bosons and Higgses, and a confined sector with one
light flavor.

In a Higgsed dark sector, the fate of the initially produced dark states
is determined by spectroscopy: they decay to lower-mass
states within the dark sector if such decays are kinematically
accessible and allowed by symmetry.  These cascades eventually produce
light states that can only decay to Standard Model final states.
Gauge boson decays are suppressed by two powers of the mixing parameter $\epsilon$, and
can be prompt or displaced.
Higgs decays can be suppressed by $\epsilon^4$, depending on
kinematics, in which case they leave the detectors before decaying to
visible matter.
Typical events in a Higgsed dark sector can produce between 4 and 12
Standard Model particles, with leptons being a significant fraction
and easiest to observe.
Caricatures of these events, with and without a recoiling
photon, are shown in Figure \ref{fig:higgsedCartoon}.
The decay phenomenology is similar in hadron colliders \cite{Strassler:2008bv,Baumgart:2009tn},
and the pure Higgsed Abelian case for B-factories has been discussed in \cite{Batell:2009yf},
though the dominant production modes and kinematics considered here are quite different.

\begin{figure}[t]
\begin{center}
\includegraphics[width=5in]{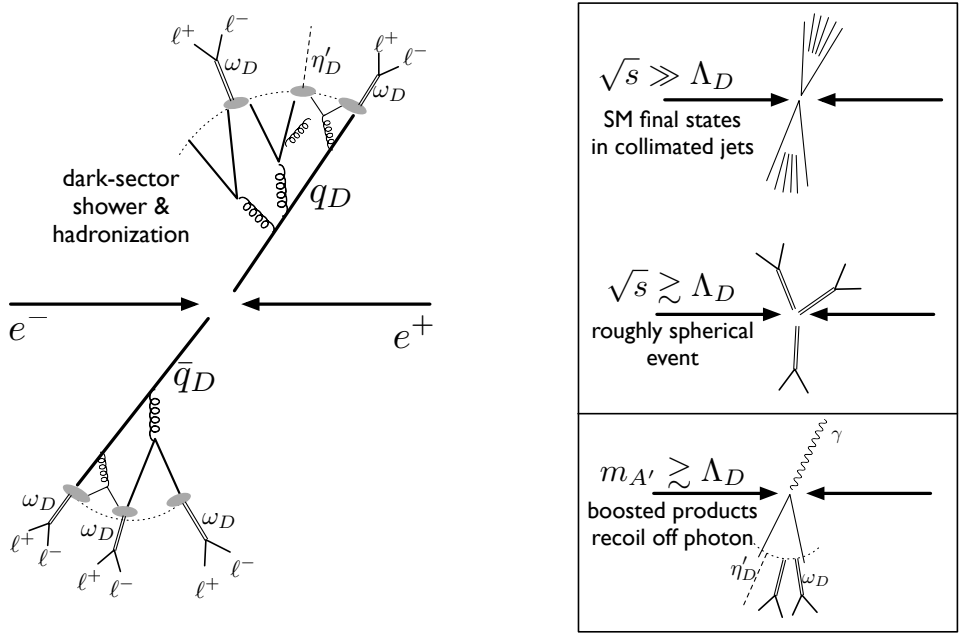}
\end{center}
\caption{Left: Cartoon of an event in which quarks in a simple
  confined dark sector are produced through an off-shell $A'$.  The
  quarks shower and hadronize into mesons, which decay into Standard
  Model particles.  Final states frequently contain many leptons, but
  can also include hadrons and long-lived dark states that escape the
  detector unobserved.  Right: Phase space structure of different
  kinds of events.  An off-shell $A'$ produces jet-like structure if
  $\Lambda_D \ll \sqrt{s}$ (top), and approximately spherical final
  states if $\Lambda_D \lesssim \sqrt{s}$ (middle).  In $A'/\gamma$
  radiative return production, the dark-sector final state recoils
  against a hard photon. \label{fig:confinedCartoon}}
\end{figure}

If the non-Abelian factor of $G_D$ is confined, then the physical
picture is very similar to the hidden valley models discussed in
\cite{Strassler:2006im,Strassler:2006qa,Han:2007ae}.  $U(1)_D$ mixing
mediates production of a light quark-antiquark pair in the dark
sector.  These states shower and hadronize, producing few dark-sector
mesons with a roughly spherical distribution if the ratio
$\sqrt{s}/\Lambda_D$ is $\OO$(1), and collimated jets if this ratio is
large.  Unlike the Higgsed scenario, the multiplicity of mesons in a
typical final state is determined by the ratio of the production
energy to $\Lambda_D$, not by spectroscopy.  Different scenarios are
caricatured in Figure \ref{fig:confinedCartoon}.  These sectors
contain light mesons that can only decay to Standard Model
final states. 
A single event can contain a combination of prompt and
displaced decays, and states that escape the detector.  In particular, 
dark pions that cannot decay within the dark sector are very long-lived because Standard Model
matter is neutral under $U(1)_D$.  This is quite different from the
decays of hidden-valley pions, mediated by a
$Z'$ under which Standard Model matter and hidden-valley matter are both
charged.  Even when all
states decay promptly, the reconstructed energy may still be less than
the collider energy if some of the many tracks in these events are not
reconstructed.  Higgsed and confined-sector decays are discussed more
carefully in Section \ref{sec:generalSector}.

These considerations motivate searching for dark-sector events in
\emph{several} channels at $e^+e^-$ colliders, as discussed in Section
\ref{sec:searches}.  We will argue that six qualitatively different
searches cover a wide range of possible phenomenology:
\begin{enumerate}
\item $4\ell$ (exclusive),
  reconstructing $E_{\rm cm}$ (also $4\ell+\gamma$)
\item $4\ell$ (exclusive), with
  displaced dilepton vertices (also $4\ell+\gamma$)
\item $\ge 5\ell + tracks$ (inclusive), reconstructing
  $E_{\rm cm}$ (also + $\gamma$)
\item $\ge 5\ell + tracks$ (inclusive), with displaced
  dilepton vertices (also + $\gamma$)
\item Very high track multiplicity, with many tracks consistent with
  leptons
\item $\gamma$ + nothing
\end{enumerate}
The combined results of searches in these regions can discover or
exclude kinetically mixed dark sectors over a wide range of mass
scales and couplings, including much of the parameter region 
motivated by inelastic dark matter. 

\section{Production Modes and Cross-Sections at $e^+e^-$ Colliders 
}\label{sec:production}

In this section, we discuss the production of dark-sector particles at
$e^+e^-$ colliders, with particular focus on the B-factories, BaBar
and BELLE.  The BaBar experiment at the SLAC National Accelerator
Center collided $e^+e^-$ pairs, obtaining 430 fb$^{-1}$ of integrated
luminosity on the $\Upsilon$(4S) resonance, at a center-of-mass energy
of 10.58 GeV.  The Belle experiment at the KEK laboratory in Japan
obtained an integrated luminosity of 725 fb$^{-1}$ on the
$\Upsilon$(4S). The two experiments also acquired $\sim$ 270 fb$^{-1}$
on the $\Upsilon$(3S) and $\Upsilon$(2S) resonances, and at nearby
energies. The $\sim 1.4$ ab$^{-1}$ of data available to the two
B-factories make these experiments ideally suited for searching for
the $\cal{O}$(GeV) dark sectors suggested by direct detection and
astrophysical data.

Over a large range of parameters, the cross-sections for the production of dark-sector particles scale as
\be\label{eqn:cross_section_scaling}
\sigma \sim \f{\alpha \alpha_D \epsilon^2}{E_{\rm cm}^2},
\ee
where $E_{\rm cm}$ is the center-of-mass energy of the collider,
$\alpha_D=\frac{g_D^2}{4\pi}$, and $g_D$ is the $A'$ gauge coupling
constant. The search sensitivity of a given $e^+e^-$ machine above
mass threshold scales as the ratio of integrated luminosity over
squared center-of-mass energy, ${\cal{L}}_{\rm int} / E_{\rm cm}^2$.  LEPI
($E_{\rm cm} \simeq 91$ GeV and ${\cal{L}}_{\rm int}\sim$ 0.5
fb$^{-1}$) and LEPII ($E_{\rm cm} \simeq$ 189-209 GeV and
${\cal{L}}_{\rm int}\sim$ 2.6 fb$^{-1}$) are much less sensitive to
direct production of low-mass dark sectors than the B-factories. As we
will review below, however, searches for rare decays of the $Z$-boson
can probe some of the relevant parameter space.

Compared to B-factories, lower-energy colliders such as DA$\Phi$NE,
which runs at $E_{\rm cm}\simeq m_\Phi \simeq 1.02$ GeV, require much
lower integrated luminosity to reach similar sensitivities to sub-GeV
dark sectors. For example, the $\sim$ 2.5 fb$^{-1}$ of data collected
by KLOE at DA$\Phi$NE should make this experiment competitive with any
B-factory searches for sub-GeV mass dark sectors, though $\sim 1/3$ as
many dark-sector events would be expected at DA$\Phi$NE than at
BELLE. Importantly, the signatures of confined dark sector models can
be quite different at $E_{\rm cm}\sim 10$ GeV compared to $E_{\rm
  cm}\sim 1$ GeV. Since there is less showering of dark sector states
for lower $E_{\rm cm}$, events at KLOE may be easier to reconstruct
than B-factory events. We will nevertheless concentrate on the
B-factory phenomenology in this work, with the understanding that at
least part of the discussion is qualitatively unchanged for several
other $e^+e^-$ colliders.

In Section \ref{subsec:production}, we quantitatively discuss the
production modes and cross-sections for general low-mass dark sectors
coupled to the Standard Model sector via kinetic mixing.  In Section
\ref{sec:DAMANormalization}, we will assume that stable matter in the
dark sector explains the direct detection signal reported by
DAMA/LIBRA, within the framework of inelastic dark matter.  Under this
assumption, we will estimate the expected production rates at
B-factories from the direct detection rate reported at DAMA/LIBRA.
Theoretically reasonable parameter ranges permit cross-sections large
enough to produce hundreds to tens of thousands of events at the
B-factories.

\subsection{Production Processes for Low-Mass Dark Sectors}\label{subsec:production}

\vskip 5mm
\begin{figure}[th]
\begin{center}
\includegraphics[width=0.9\textwidth]{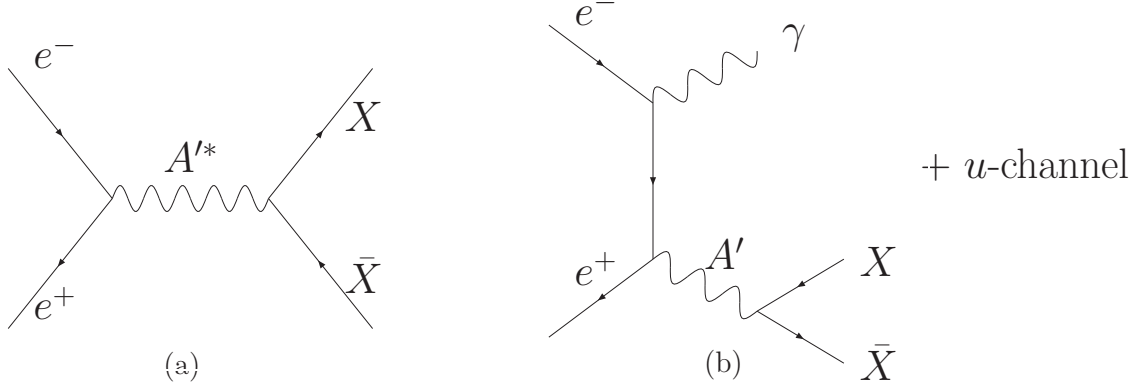}
\caption{\label{fig:diagrams} Production modes of light dark-sector particles $X\bar{X}$ at B-factories.
Left: Production through an off-shell $A'$.
Right:  Production of an on-shell $A'$ and a photon --- we assume the $A'$ subsequently decays into lighter
dark-sector particles.}
\end{center}
\end{figure}

At $e^+e^-$ colliders, there are two important dark-sector production
modes.  For definiteness, we focus here on the production
of a pair of dark fermions, but the production cross-section for a pair of
dark gauge bosons in a Higgsed dark sector, which occurs through mass
mixing of the $A'$ with non-Abelian gauge bosons, is similar.

We first consider the direct production of a pair $X\bar X$ of dark-sector fields through an off-shell
$A'$, shown in Figure \ref{fig:diagrams}(a).
The cross-section for this off-shell production process is
\bea
\sigma_{X\bar X} & = & N_c\, \frac{4\pi}{3}\,\frac{\epsilon^2 \alpha \alpha_D}{E_{\rm cm}^2}
\,\Bigg|1-\frac{m_{A'}^2}{E_{\rm cm}^2}-\frac{i m_{A'}\Gamma}{E_{\rm cm}^2}\Bigg|^{-2}
\sum_{i=1}^{N_f}\,q_i^2\,\sqrt{1-\f{4m_{X_i}^2}{E_{\rm cm}^2}}\, \left(1+\f{2 m_{X_i}^2}{E_{\rm cm}^2}\right)  \nonumber \\
& \simeq &
0.78\;{\rm fb}\;N_c\,\left(\f{\epsilon}{10^{-3}}\right)^2\; \left(\f{\alpha_D}{\alpha}\right)\; \left(\f{E_{\rm cm}}{10.58\mbox{ GeV}}\right)^{-2} \times N_f.
\label{eq:off-shell}
\eea
Here, $N_c$ is the number of colors in the dark sector gauge group $G_D$, $N_f$ is the number of
dark sector particles $X_i$ coupling to $A'$ with charges $q_i$, and the other variables
were defined in equation (\ref{eqn:cross_section_scaling}).
This cross-section is non-zero even for $m_{A'}$ much larger than $E_{\rm cm}$, as long as
$m_{X} < E_{\rm cm}/2$.

\begin{figure}[t]
\includegraphics[width=.48\textwidth]{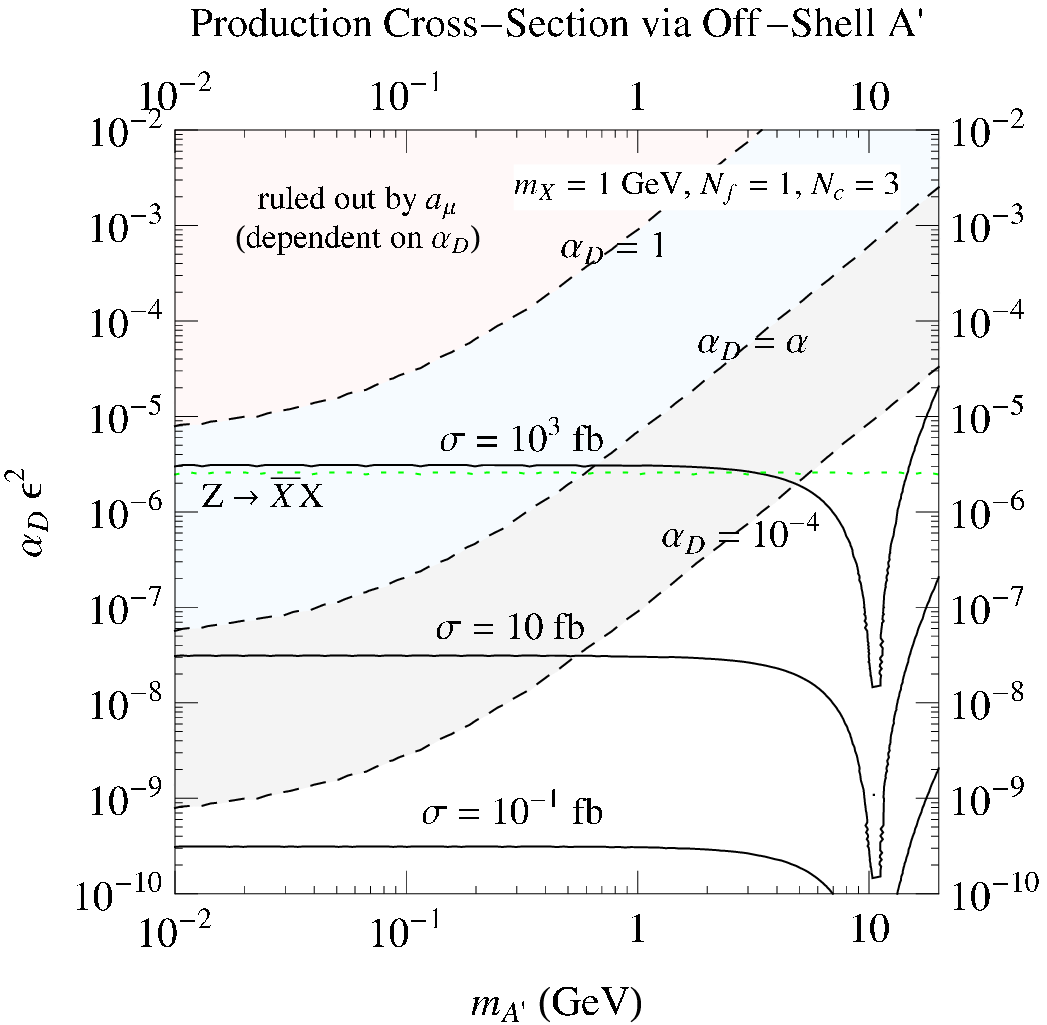}
\includegraphics[width=.48\textwidth]{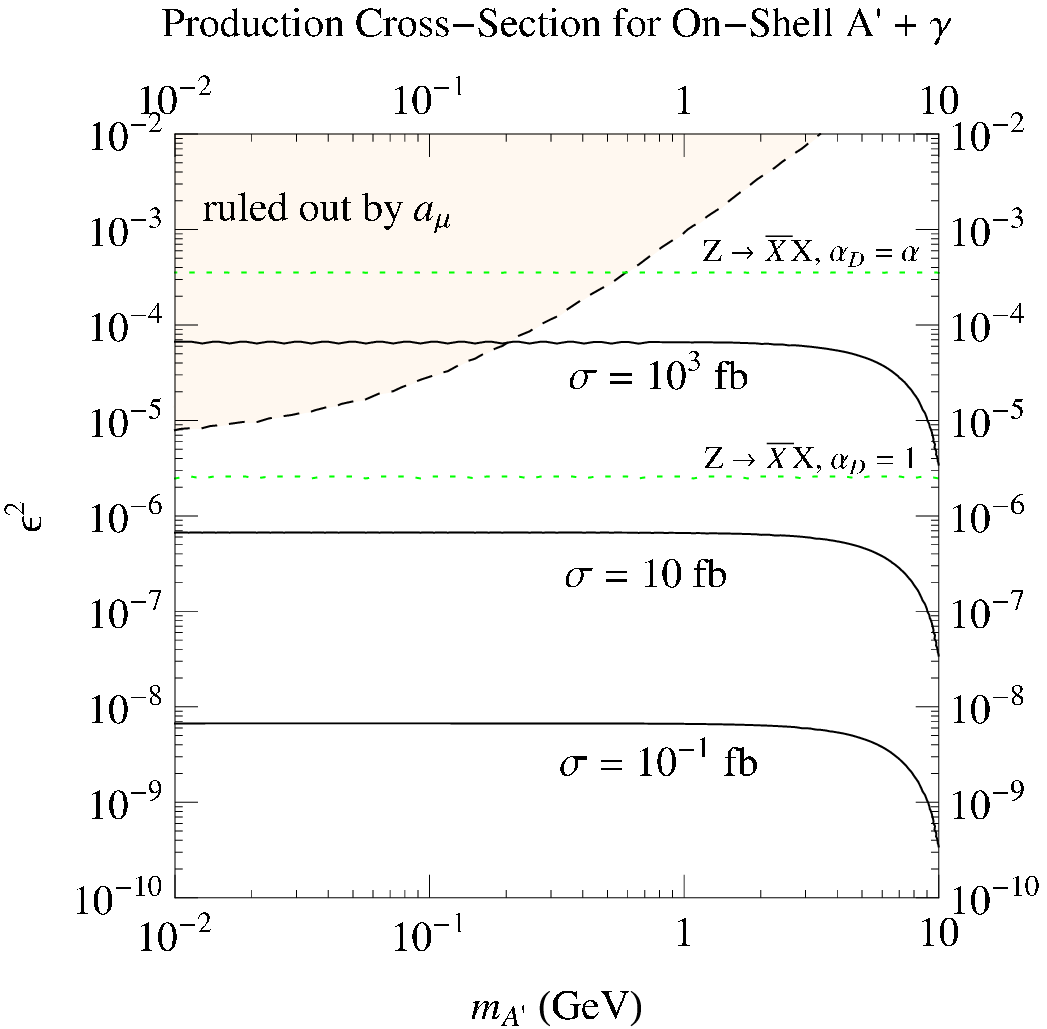}
\caption{\label{fig:BfactoryCrossSection} 
Left: Inclusive cross-section at the B-factories $(E_{\rm cm}=10.58\mbox{ GeV})$ for production of dark-sector states, $X$, via 
an off-shell $A'$ as a function of $\alpha_D\epsilon^2$ and $m_{A'}$, for $m_X=1$ GeV, $N_f=1$, 
$N_c=3$, and unit charges $q_i$ (equation \eqref{eq:off-shell}). 
Note that the cross-section scales linearly with the number of dark flavors $N_f$ and dark colors $N_c$.
Right: Cross-section at the B-factories for production of an on-shell $A'$ and a photon as a function of 
$\epsilon^2$ and $m_{A'}$ (equation \eqref{eq:onshellXsectionnoapprox}).
Black lines correspond to fixed cross-sections.
Also shown are the constraints on the couplings of a new $U(1)_D$ mixing with hypercharge from measurements of
the muon anomalous magnetic dipole moment (shaded regions) \cite{Pospelov:2008zw}.
The green dotted lines correspond to the lower bounds on the range of couplings that could be probed
by a search at LEPI for rare $Z$-decays to various exotic final states, assuming that branching ratios as low
as $10^{-5}$ can be probed.  More details are given in Appendix \ref{app:Contraints}.}
\end{figure}

\begin{figure}[t]
\includegraphics[width=.48\textwidth]{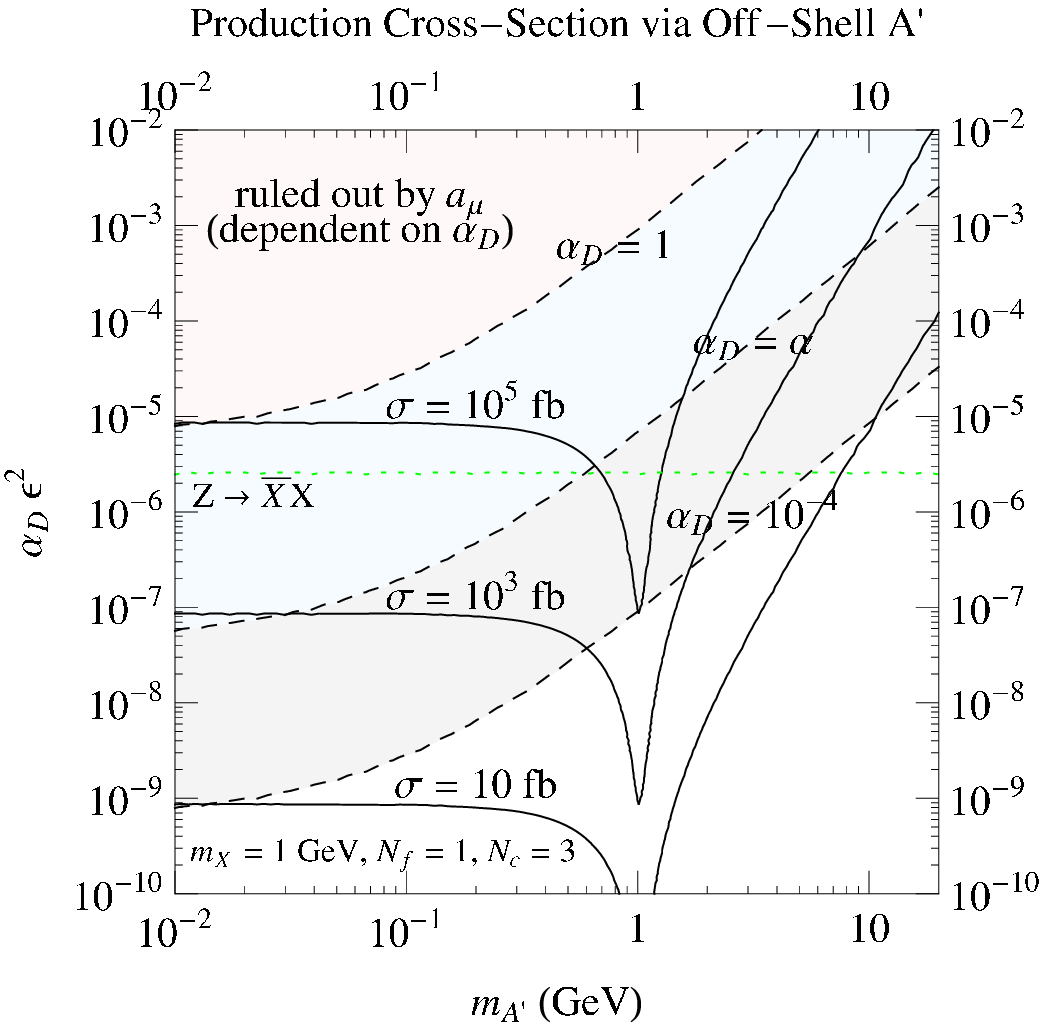}
\includegraphics[width=.48\textwidth]{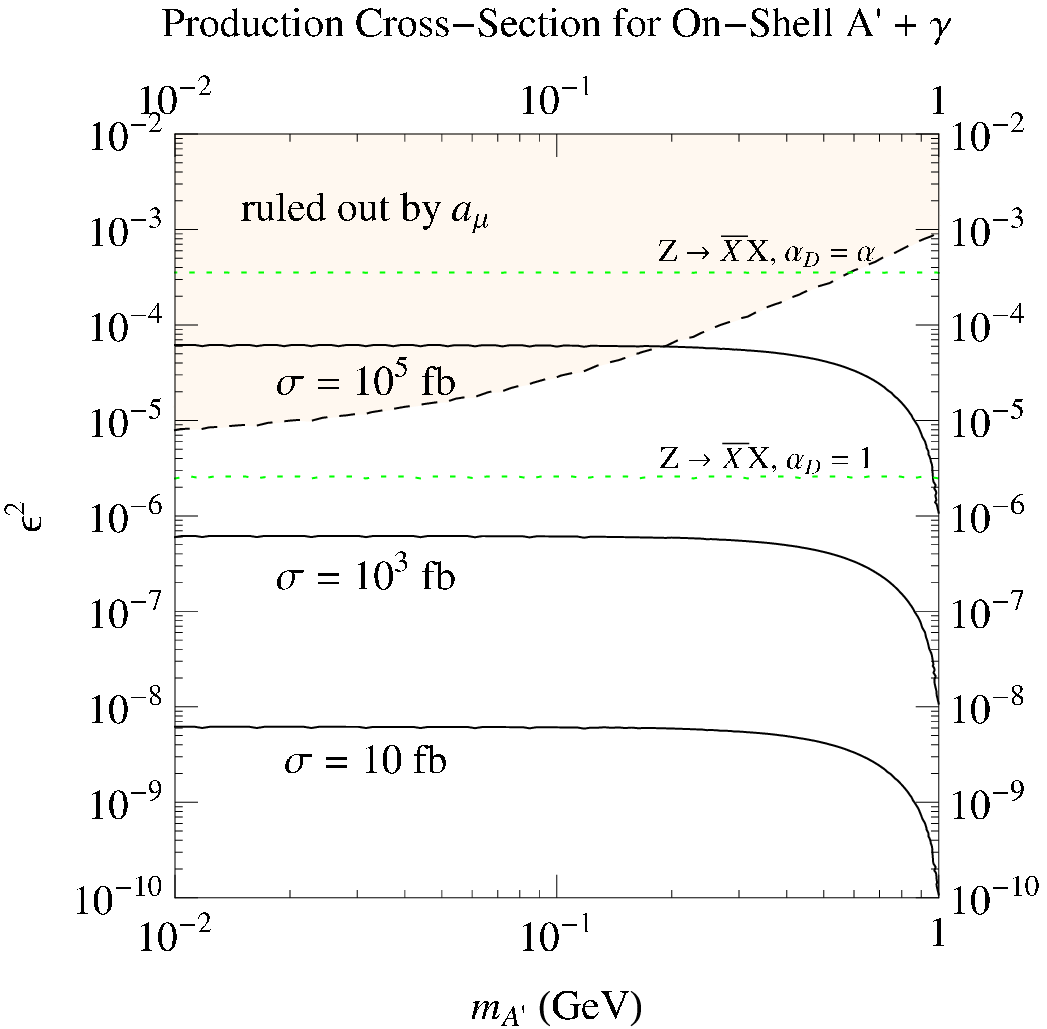}
\caption{\label{fig:DaphneCrossSection} As in Figure \ref{fig:BfactoryCrossSection}, but for the center-of-mass energy of
DA$\Phi$NE $(E_{\rm cm}=1.02\mbox{ GeV})$.}
\end{figure}

The second production mode, shown in Figure \ref{fig:diagrams}(b), is that of an on-shell $A'$
and a photon, with differential cross-section
\be
\f{d\sigma_{\gamma A'}}{d\cos\theta} =
\f{2\pi \epsilon^2 \alpha^2}{E_{\rm cm}^2}\left(1-\f{m_{A'}^2}{E_{\rm cm}^2}\right)
\f{1+\cos^2\theta + \f{4m_{A'}^2/E_{\rm cm}^2}{(1-m_{A'}^2/E_{\rm cm}^2)^2}}{(1+\cos\theta)(1-\cos\theta)},
\ee
where $\theta$ is the angle between the beam line and the photon momentum.
This has singularities as the photon becomes collinear with the
initial-state electron or positron. The singularity is cut off by the electron mass, but if we demand that the photon be
detected in the electromagnetic calorimeter in the range
$\cos\theta_{\rm min} < \cos\theta < \cos\theta_{\rm max}$, the
cross-section is  
\bea
\sigma & = & \f{2\pi\epsilon^2\alpha^2 }{E_{\rm cm}^2}\left(1-\f{m_{A'}^2}{E_{\rm cm}^2}\right) \left(\Big(1+\f{2m_{A'}^2/E_{\rm cm}^2}{(1-m_{A'}^2/E_{\rm cm}^2)^2}\Big) \Theta -\cos\theta_{\rm max} + \cos\theta_{\rm min} \right) \label{eq:onshellXsectionnoapprox} \\ 
& \sim & 2.4\,{\rm fb}\, \Big(\f{\epsilon}{10^{-3}}\Big)^2 \Big(\f{E_{\rm cm}}{10.58}\Big)^{-2}
\Big(\log\f{4}{\theta_{\rm min}(\pi-\theta_{\rm max})}- 1 \Big), \label{eq:onshellXsection}
\eea
where
\bea
\Theta \equiv \log\left(\f{(1+\cos\theta_{\rm max})(1-\cos\theta_{\rm min})}{(1+\cos\theta_{\rm min})(1-\cos\theta_{\rm max})}\right) \approx 6 
\eea
is an enhancement from the $t$-channel singularity (the numerical
estimate is obtained from the BaBar calorimeter range in the center-of-mass
frame, $-0.92 \lesssim \cos(\theta_{min}) \lesssim 0.87$ \cite{Harrison:1998yr}), and we
assume $m_{A'}^2 \ll E_{\rm cm}^2$ in equation (\ref{eq:onshellXsection}).  The
on-shell process (\ref{eq:onshellXsection}) is enhanced by $\sim \f{\alpha}{\alpha_D}
\Theta$ relative to the off-shell mode (\ref{eq:off-shell}).

For $m_{A'}\sim 1$ GeV, $\alpha_D = \alpha$, and $\epsilon =10^{-3}$,
the relevant cross-sections are $\OO$(fb) at the B-factories. With a
total integrated luminosity of $\sim 1.4$ ab$^{-1}$, this corresponds
to several hundred to a few thousand dark-sector production events!
In Figure \ref{fig:BfactoryCrossSection}, we show the inclusive
cross-sections at the B-factories for off-shell and on-shell
production of dark-sector particles as a function of either
$\alpha_D\epsilon^2$ or $\epsilon^2$ and $m_{A'}$.  Figure
\ref{fig:DaphneCrossSection} shows a similar plot for DA$\Phi$NE.

Figures \ref{fig:BfactoryCrossSection} and
\ref{fig:DaphneCrossSection} also show constraints on the couplings
of a new $U(1)_D$ mixing with hypercharge. Shown in shaded regions are
constraints derived from the muon anomalous magnetic dipole moment
$a_{\mu}$ as in \cite{Pospelov:2008zw}.  The measured value of $a_{\mu}$ is
larger than the theoretically predicted value, based on $e^+e^-$
annihilation to hadrons, by $(302 \pm 88) \times 10^{-11}$ \cite{Passera:2008hj,Bennett:2006fi},
and the constraints shown assume
that the contribution from loops of the $A'$ to $a_{\mu}$ is smaller
than $(302+5\times 88) \times 10^{-11}$.
This constraint is discussed in more detail in Appendix \ref{app:Contraints}.

Since any mixing between $U(1)_D$ and $U(1)_Y$ induces a coupling
between $Z$-bosons and matter in the dark sector, decays $Z\to
X\bar{X}$ are possible. Depending on the decay modes of $X$, which we
will discuss in detail in Section \ref{sec:generalSector}, a variety of
signatures are possible.  For example, if the $X$'s decay invisibly or
outside the detector, the constraint is not better than the
uncertainty on the width of the $Z$, which is about $0.1\%$.  More
exotic $Z$ decays into multi-leptons are possible, but require
dedicated searches.  The best constraints on rare $Z$ decay branching
ratios are no better than the $10^{-6}-10^{-5}$ level \cite{Amsler:2008zzb}, and
even dedicated searches may not reach this level for complicated final
states.  In Figures \ref{fig:BfactoryCrossSection} and \ref{fig:DaphneCrossSection},
we show (green dotted lines) the value of $\alpha_D\epsilon^2$ corresponding to a
branching ratio of $Z\to X\bar{X}$ as small as $10^{-5}$.  The
potential reach of rare $Z$-decay searches are discussed in more detail in Appendix \ref{app:Contraints}.

Finally, we mention two collider constraints on the production cross-section
which are only relevant for certain decay modes.
The first one is from the CLEO collaboration which searched for decays `$\Upsilon(1S)\to \gamma$ + nothing' \cite{Balest:1994ch}.
This is relevant for the case when the $X$'s are long-lived and decay outside
the detector.
As we will see in Sections \ref{sec:generalSector} and \ref{sec:searches}, this is
a very important search mode.
Using an integrated luminosity of 48 pb$^{-1}$ of data, CLEO found no
peak in the photon energy in the range 1-4.7 GeV
\cite{Balest:1994ch}.  The resulting limit can be
interpreted as excluding cross-sections as low as 200-500 fb, depending on
the $A'$ mass, though \cite{Balest:1994ch} assumes a different photon
angular distribution than would result from dark-sector production.

The second collider constraint comes from a search for $\Upsilon(3S)
\to \gamma A' \to \gamma \mu^+ \mu^-$ by the BaBar collaboration
\cite{Aubert:2009cp}, which is only relevant if $A'$ decays to
dark-sector particles are kinematically forbidden, in which case it
decays to Standard Model states.  In 30 fb$^{-1}$ of data containing
$\sim 122\times 10^6$ $\Upsilon(3S)$ events, a 90\% C.L.~upper limit
of roughly $4\times 10^{-6}$ on the $\gamma \mu^+\mu^-$ branching
fraction was found for $m_{A'}$ of a few hundred MeV.  This search
would also be sensitive to about 500 $\gamma\mu^+\mu^-$ events through
$A'/\gamma$ production.  In this mass range,
$Br(A'\to\mu^+\mu^-)\approx 0.3-0.5$ \cite{Amsler:2008zzb}.  These results exclude
production of about 1000 $\gamma A'$ events, corresponding to a
cross-section of about 30 fb.  For $A'$ masses of a few GeV, but below
the $\tau$ threshold, the limits exclude 250 $\gamma\mu^+\mu^-$
events, but $Br(A'\to\mu\mu) \approx 0.2-0.25$ so that a similar
cross-section limit is obtained. We note that even if the $A'$ is the
lightest state of the dark sector, any other kinematically accessible
states are pair-produced with a comparable rate through an off-shell
$A'$, which produces a signal in the multi-lepton channel (see Section
\ref{sec:generalSector}), which has much lower QED backgrounds.  Lower
cross-sections can thus be probed by considering different final
states.

In summary, the potential search mode for rare $Z$ decays and the
$a_{\mu}$ constraint do not exclude parameter regions in which
thousands of events could be seen.  Collider searches for
`$\gamma+\mbox{nothing}$' and $\gamma+\mu^+\mu^-$ final states have
probed some of this parameter space, but are quite model-dependent.
More general final states could easily have been missed by existing
searches, even if they occur with large cross-sections.

Before discussing the decay of $X$'s in Section
\ref{sec:generalSector}, we use the hypothesis that an
$\mathcal{O}$(100 GeV) stable particle charged under $G_D$ explains
the direct detection signal reported by DAMA/LIBRA.
We use this to estimate the dark sector couplings
$\alpha_D\epsilon^2$ and resulting production cross-sections of
low-mass dark sector states at B-factories.

\subsection{Inelastic Dark Matter and Inclusive Cross-Sections}\label{sec:DAMANormalization}
Inelastic dark matter (iDM) can reconcile the annual
modulation signal reported by the DAMA/LIBRA collaboration with the null results of other
experiments \cite{TuckerSmith:2001hy,Chang:2008gd}.
In proposals to generate the iDM splitting from new
non-Abelian gauge interactions in a $\sim$ GeV-mass dark sector, it is
also natural to assume that scattering is mediated by a $U(1)_D$ gauge
boson $A'$ that mixes kinetically with the photon, with mass $m_{A'}$
of $\OO$(100 MeV - 10 GeV) \cite{ArkaniHamed:2008qn}.
In such models, the signal rate
and spectrum reported by DAMA/LIBRA constrain the couplings of the $A'$
to the Standard Model electromagnetic current.
This allows us to estimate the inclusive cross-section for production of
dark states at B-factories for a given iDM model.

At energies beneath $m_{A'}$, the dark matter interacts with charged
matter through an effective coupling
\bea
\mathcal{L} \simeq
\frac{eg_D\epsilon}{m_{A'}^2}J^{\mu}_{EM}J_{\mu,\text{Dark}} \label{eq:InteractionLagrangian2},
\eea
where $J_{\mu,\text{Dark}}$ is a current composed of dark sector
fields, and $e$ and $g_D$ are the photon and $A'$ coupling strengths,
respectively.
The combination $\frac{\alpha_D\epsilon^2}{m_{A'}^4}$
($\alpha_D = g_D^2/4\pi$) controls the scattering rate of dark matter
in direct detection experiments (see Appendix \ref{app:iDMreview}),
while $\alpha_D \epsilon^2$
determines the inclusive production cross-section at B-factories.
Consequently, the production cross-section can be estimated as a function of $m_{A'}$.

\begin{figure}[t]
\includegraphics[height=2.5in,width=3.25in]{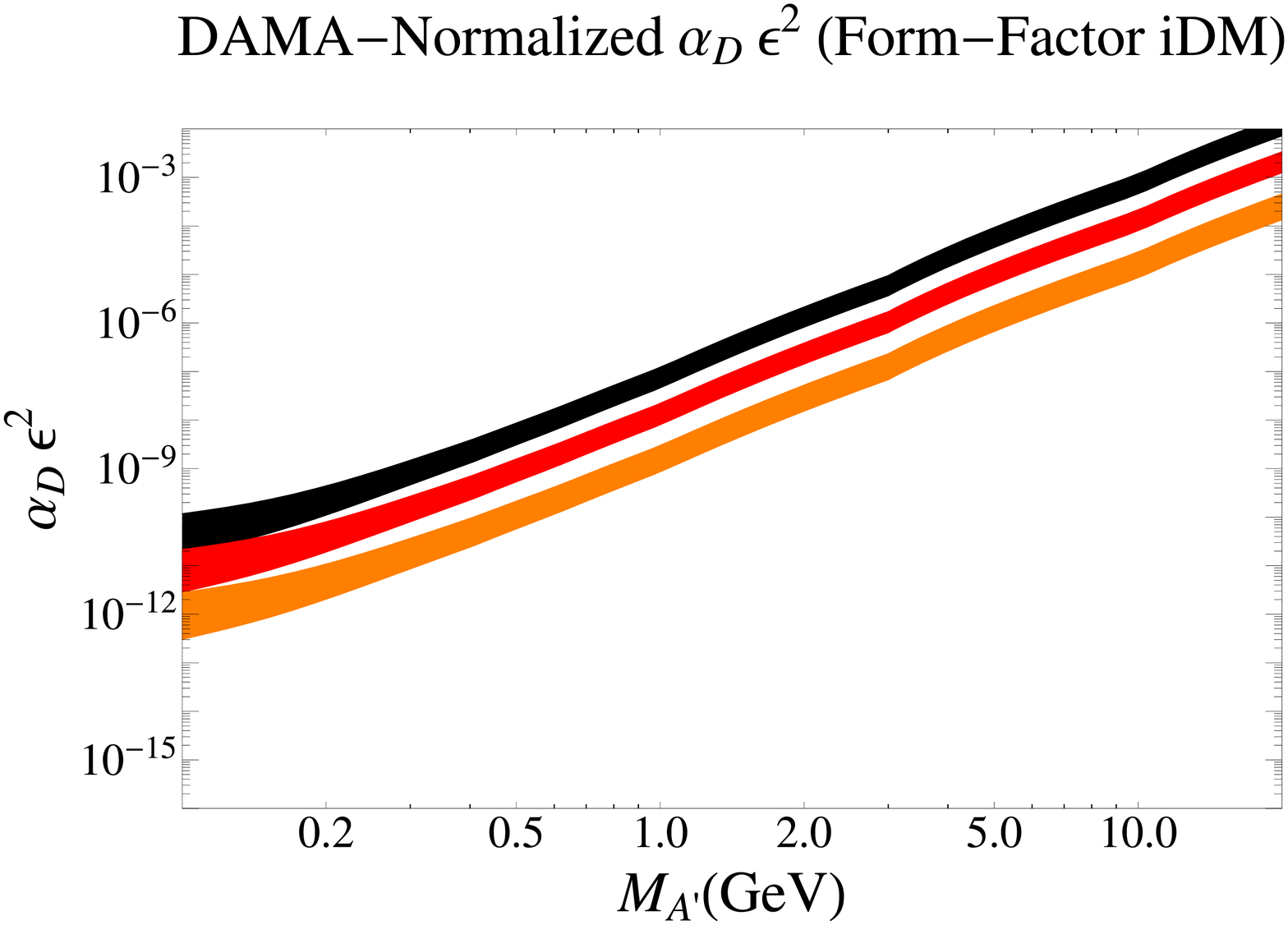}
\includegraphics[height=2.5in,width=3.25in]{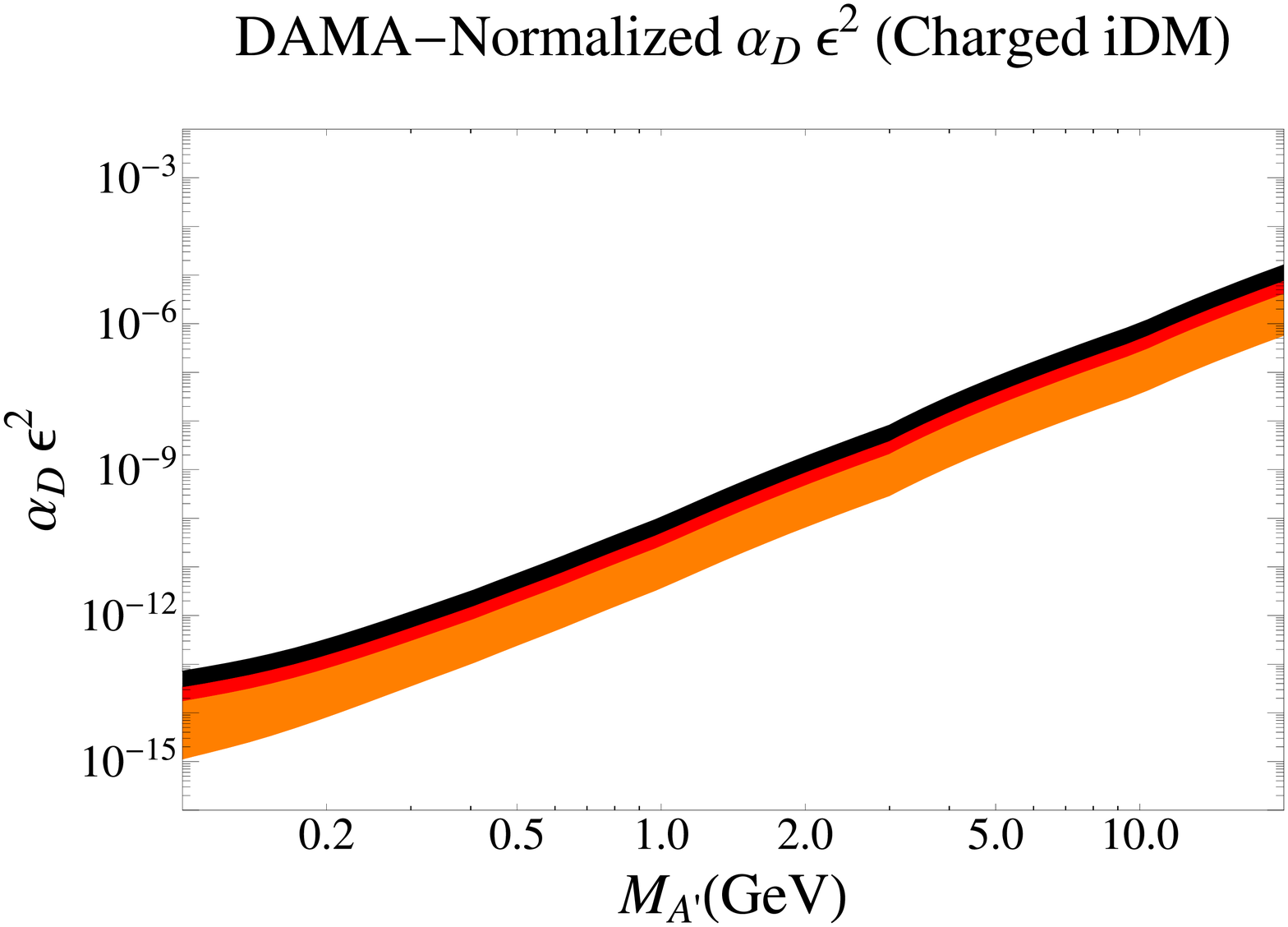}
\caption{\label{fig:DAMANormCoupling} $\alpha_D\epsilon^2$ as a
  function of $m_{A'}$, normalized to the DAMA/LIBRA annual modulation
  signal. Black, red, and orange bands from top to bottom correspond
  to $m_{DM}=2000, 800$, and 200 GeV, 
  respectively. Uncertainties in fitting to DAMA/LIBRA, as a function
  of the inelastic dark matter mass splitting, are reflected in the
  thickness of the bands. The left plot shows the normalized couplings
  for inelastic dark matter that scatters via form-factor-suppressed
  dipole scattering (equation \eqref{eq:Dipole}), typical for a
  confined dark sector, with $\tilde C=0.01$ ($\alpha_D\epsilon^2$
  scales as $1/\tilde C$). The right plot shows the normalized
  couplings for charged scattering (equation \eqref{eq:VectorScalar}),
  which is typical for a Higgsed dark sector.}
\end{figure}
\begin{figure}[t]
\includegraphics[height=2.5in,width=3.25in]{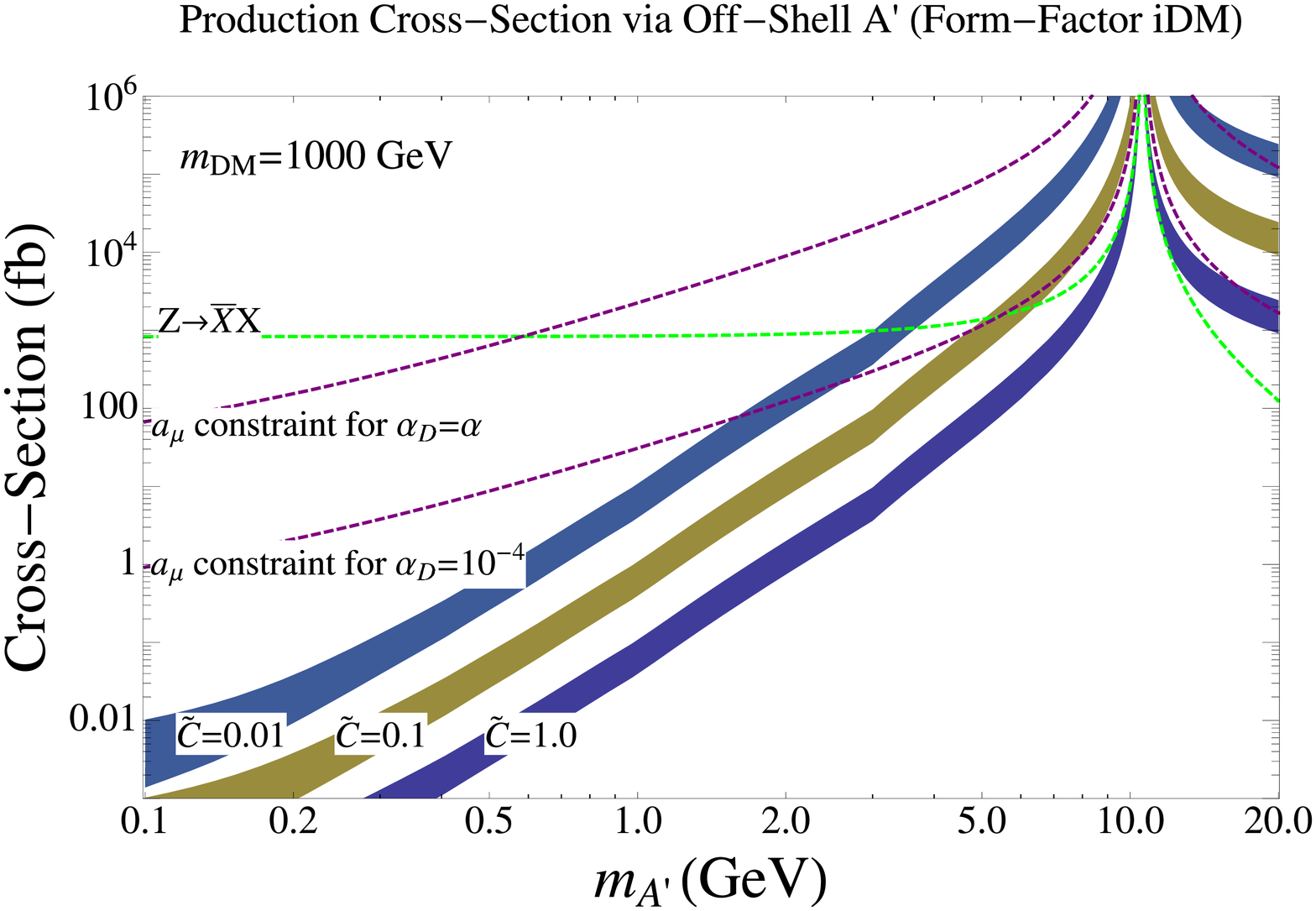}
\includegraphics[height=2.5in,width=3.25in]{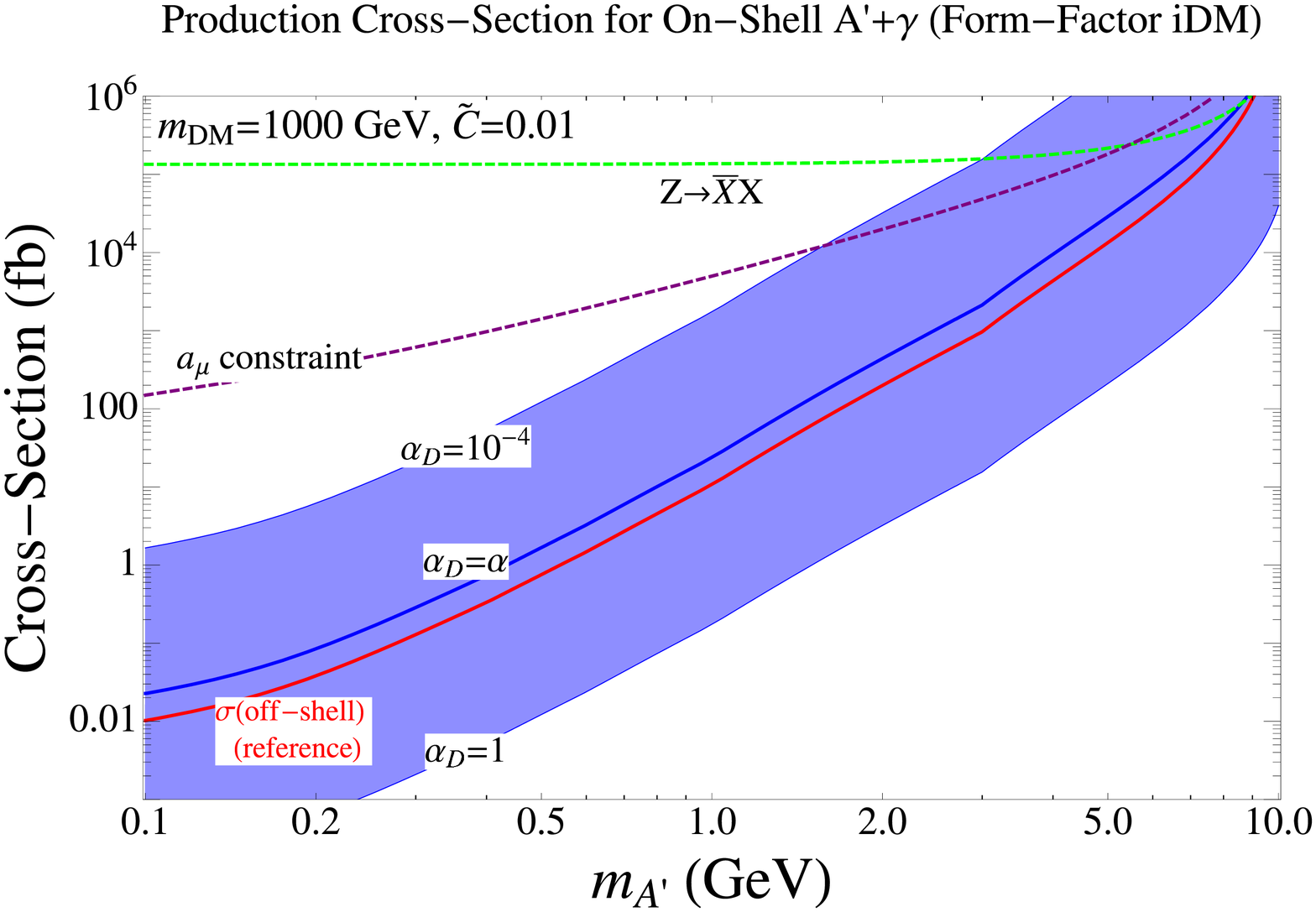}
\caption{\label{fig:CompositeCrossSections} Inclusive cross-section at
  the B-factories for production of dark-sector states, $X$, through
  an off-shell $A'$ (left) and for production of an on-shell $A'$ with
  a photon (right) as a function of $m_{A'}$, after normalizing to the
  observed DAMA/LIBRA modulation rate assuming form-factor-suppressed
  dipole inelastic dark matter scattering.  We set $m_{DM}=1$ TeV.  In
  the left plot, we set $m_X=1$ GeV, $N_f=1$, $N_c=3$, $q_i=1$, and
  show various choices of $\tilde{C}=\{0.01,0.1,1.0\}$, while in the
  right plot, we show $\alpha_D=\{1,\alpha,10^{-4}\}$ with
  $\tilde{C}=0.01$.  Note that the off-shell production cross-section
  scales linearly with the number of dark flavors $N_f$ and dark
  colors $N_c$.  Also shown on both plots are constraints from
  $a_{\mu}$ as well as the rare $Z$ decay sensitivity region --- see
  Appendix \ref{app:Contraints} for more details.}
\end{figure}
\begin{figure}[t]
\begin{center}
\includegraphics[height=2.5in,width=3.3in]{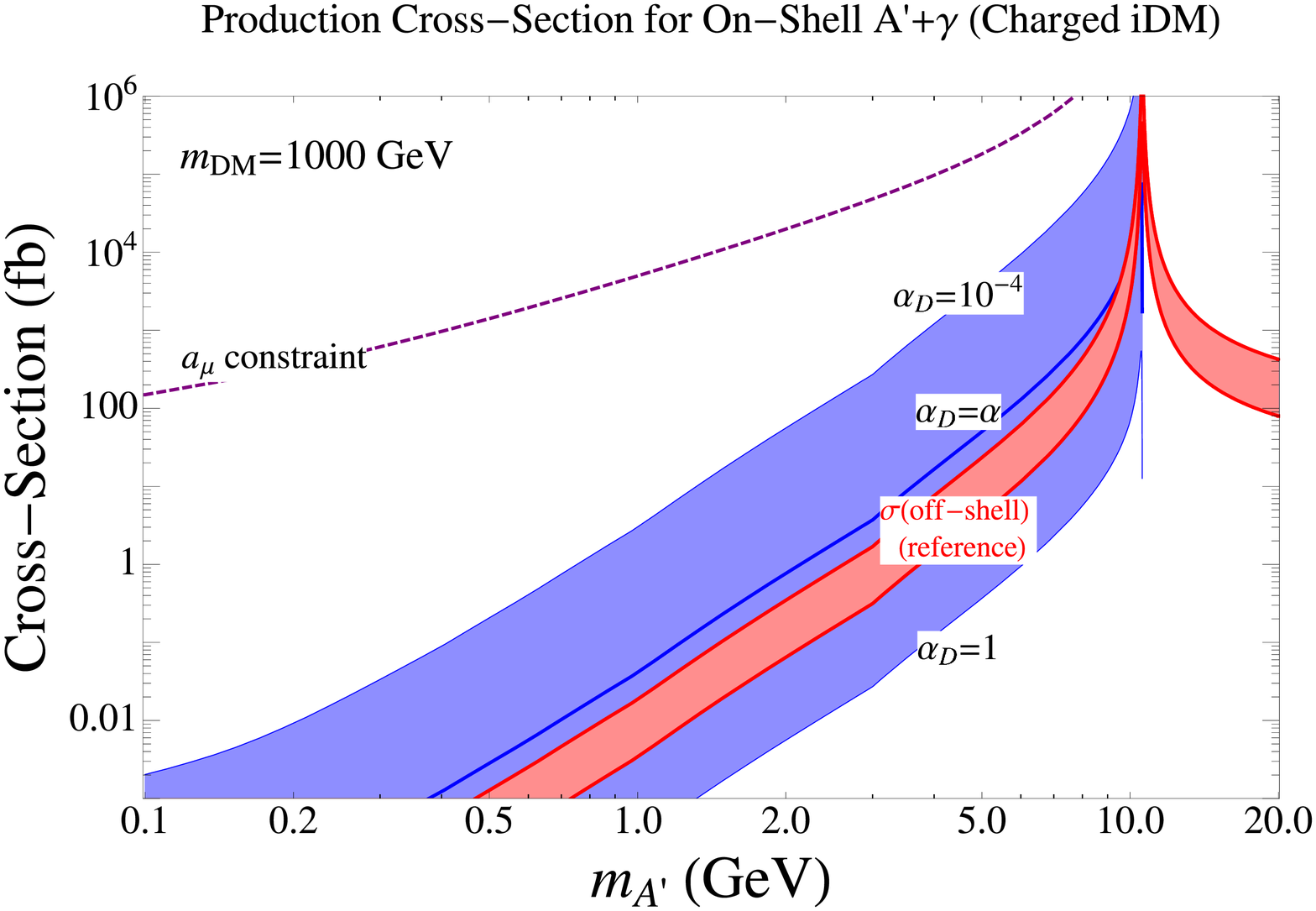}
\caption{\label{fig:ChargedCrossSections} As in Figure \ref{fig:CompositeCrossSections}, but here the production cross-sections 
are normalized to the observed DAMA/LIBRA modulation rate assuming charged inelastic dark matter scattering.}
\end{center}
\end{figure}

We will consider two classes of iDM theories --- one where the dark
matter is directly charged under the $U(1)_D$
\cite{ArkaniHamed:2008qn}, and one where the dark matter is a neutral
composite meson, built out of charged spin-1/2 constituents, one of
which is light ($\sim$GeV) and the other heavy ($m_{DM}\sim 100-1000$ GeV)
\cite{Schuster:2009}.  Higgsed dark sectors are an example of the
first class of theories, in which the breaking of the dark gauge group
generates mass splittings for the dark matter gauge eigenstates.  The
resulting mass eigenstates couple off-diagonally to the $A'$, leading
to inelastic scattering.  The low-energy scattering matrix elements
are
\bea
\langle \chi^*(p')|J^{\mu}_{\text{Dark}}|\chi(p)\rangle_{\text{Charged Scattering}} &=& (p+p')^{\mu} ,\label{eq:VectorScalar}
\eea
where we have neglected higher order spin-dependent terms.
Confined dark sectors are an example of the second class of theories, in which
hyperfine interactions among the constituents
split the ground state meson into a pseudo-scalar and vector \cite{Schuster:2009}.
If the $J^{\mu}_{\text{Dark}}$ coupling breaks parity, then the leading scattering operators
mediate inelastic hyperfine transitions (dipole scattering) with
matrix elements
\bea
\langle \chi_V(p',\epsilon^\mu)|J^{\mu}_{\text{Dark}}|\chi_S(p) \rangle_{\text{Dipole Scattering}} &\approx& \frac{c_{\text{in}}}{\Lambda_D}\left(p+p'\right)^\mu\epsilon\cdot q,\label{eq:Dipole}
\eea
where $\epsilon^{\mu}$ is the polarization of the outgoing $\chi_V$
and $q=p-p'$ is the momentum transfer.
The natural magnitude of the dipole moment $b=\frac{c_{\text{in}}}{\Lambda_D} \sim$ GeV$^{-1}$
is the size $\Lambda_D^{-1}$ of the composite dark matter
($\Lambda_D$ is of order the confining scale of the dark gauge
group).  Relative to the charged scattering matrix element
\eqref{eq:VectorScalar}, this scattering is suppressed by a
$q$-dependent form factor.

For either current, the cross-section to scatter off of a nuclear target of mass $m_N$ and charge $Z$ with recoil energy $E_R$ is,
\be
\frac{d\sigma}{dE_R} \approx \frac{8\pi Z^2\alpha \alpha_D \epsilon^2 m_N}{v^2(2 m_N E_R+ m_{A'}^2)^2}|F(E_R)|^2
\begin{cases}
1 & (\text{charged (vector-current)}) \vspace{.5cm}\\
\frac{c_{\text in}^2}{2}\frac{m_N\,E_R}{\Lambda_D^2} & (\text{form-factor (dipole)})
\end{cases}\label{eq:dSdER}
\ee
where $\alpha=\frac{e^2}{4\pi}$, $v$ is the dark matter-nucleus
relative velocity, and $F(E_R)$ is a nuclear form factor.
As discussed in Appendix \ref{app:iDMreview}, we compute the scattering rate from
equations (\ref{eq:VectorScalar}) and (\ref{eq:Dipole}), fit the reported DAMA/LIBRA rate,
and thereby obtain an estimate for the couplings $\alpha_D\epsilon^2$
as a function of $m_{A'}$.

From equation (\ref{eq:dSdER}), we see that the dipole scattering cross-section is a function of the
momentum-dependent form factor $\frac{m_N E_R}{\Lambda_D^2}\sim \frac{q^2}{\Lambda_D^2}$.
To estimate the coupling $\alpha_D\epsilon^2$ for a given $m_{A'}$, we must thus choose a value for $\Lambda_D$.
$\Lambda_D$ can in principle be determined by fitting $\delta$ to the DAMA/LIBRA data, since $\delta$ is
proportional to $\Lambda_D^2/{m_{DM}}$.
However, the constant of proportionality is model-dependent.
Following \cite{Schuster:2009}, we introduce a scale $\tilde{\Lambda}_D$,
such that
\bea
\delta &\equiv& \frac{\tilde{\Lambda}_D^2}{m_{DM}}.
\eea
In general,
\be
\tilde{\Lambda}_D \sim
\begin{cases}
\frac{\Lambda_D}{\sqrt{N_c}}  & (\text{strong coupling})\\
\alpha'\Lambda_D & (\text{coulomb regime})
\end{cases}
\ee
where the first case corresponds to a heavy meson of a $SU(N_c)$
confining at $\Lambda_D$, and the second case corresponds to a meson
in a weak-coupling coulomb regime with coupling $\alpha'$.
$\tilde{\Lambda}_D/\Lambda_D$ is naturally a small number in either
case.  We absorb the uncertainty in both $\Lambda_D$ and
$c_{\text in}$ in a single proportionality factor 
\bea
\tilde{C}\equiv \left(\frac{c_{\text in}^2\tilde{\Lambda}_D^2}{\Lambda_D^2}\right),
\eea which is naturally small ($\tilde C \sim 0.01-1$).  The dark
matter scattering rate is now exactly fixed by the inelastic splitting
$\delta$ and $\tilde C$, allowing the determination of $\alpha_D\epsilon^2$
from the DAMA/LIBRA scattering rate as a function of $m_{A'}$.  For
the charged scattering \eqref{eq:VectorScalar}, there is no analogous
unknown parameter, and we can solve for $\alpha_D \epsilon^2$ directly
from \eqref{eq:dSdER}.

Figure \ref{fig:DAMANormCoupling} shows the DAMA/LIBRA-normalized couplings
as a function of $m_{A'}$ for charged and dipole
scattering, and for different dark matter masses.
In Figure
\ref{fig:CompositeCrossSections}, we show the B-factory cross-sections
using dark sector couplings normalized for dipole scattering.
In Figure \ref{fig:ChargedCrossSections}, we show cross-sections
normalized using charged scattering.
For $m_{A'}\geq 100$
MeV, B factory cross-sections are expected to easily exceed $\sim 1$
fb for composite iDM theories. For theories where the iDM scattering
is charged-current dominated, $\sim 1$ fb cross-section can occur for
$m_{A'}\geq 1$ GeV.
In general, the estimated cross-sections scale as $m_{A'}^4$.

We note that the $e^+$ and/or $e^-$ data from, for example, 
PAMELA, ATIC, and PPB-BETS cannot be used in the same way 
as the DAMA data to predict the cross-sections at low-energy 
$e^+e^-$ colliders.  The reason is that the $e^+$ and/or $e^-$ 
data 
does not directly constrain the $A'$ coupling with the electromagnetic 
current.  
Since the dark matter annihilates into \emph{on-shell} $A'$s that decay 
with a lifetime that is largely unconstrained by the data, the mixing 
parameter $\epsilon$ is also unconstrained.  
However, indirect constraints on the dark matter annihilation cross-section 
from, for example, $\gamma$-ray observations \cite{Kamionkowski:2008gj,Essig:2009jx} and 
CMB measurements \cite{Galli:2009zc} constrain the maximum allowable Sommerfeld 
enhancement, and currently disfavor very low $A'$ masses ($\sim$ 100 MeV or less).

\section{Decays of Dark-Sector States into Standard Model Particles}\label{sec:generalSector}
In this lengthy section, we discuss the decays of dark-sector states
produced at $e^+e^-$ colliders.  Readers interested in the
experimental consequences of these decays can skip to Section
\ref{sec:searches}, which is self-contained.

The kinetic mixing that allows dark-sector states to be produced at
$e^+e^-$ colliders also allows them to decay back into Standard Model
states, mainly leptons and light hadrons.  To demonstrate the range of
phenomenology possible in both Higgsed and confined dark sectors, we
have considered several kinematic limits that exhibit different
properties.
We will also briefly consider the impact of Higgs mixing on decays.
We organize our discussion around the following questions:
\begin{enumerate}
\item What are the metastable states of the dark sector that can
  \emph{only} decay through processes that involve kinetic
  mixing?
\item What are the decay widths of these metastable states?
\item What are the multiplicities and kinematic properties of these
  metastable states?
\end{enumerate}

We consider Higgsed dark sectors in Section \ref{subsec:higgsedDS} and
confined dark sectors in Section \ref{subsec:confinedDS}.  In both
cases, decays of spin-1 states through $A'$ mixing, suppressed by
$\epsilon^2$, are generically prompt, while other species can have
displaced 3-body decays or invisible decays suppressed by
$\epsilon^4$.  Therefore, different dark-sector spectra can give rise
to very different phenomenology, with all dark-sector states decaying
promptly into the standard model, all states so long-lived that they
escape the detector, or a combination of prompt, displaced, and/or
invisible decays.  Light fermions in the dark sector lead to events
with more invisible decay products, as we discuss in Section
\ref{subsec:fermions}.

Through most of this discussion we consider only the decays generated
by the $A'/\gamma$ kinetic mixing, but other interactions between the
Standard Model and dark sector, involving either Standard Model Higgs
bosons or new heavy fields coupled to both sectors, are also possible.
Though these have no observable effect on direct production, they can
dominate over the kinetic-mixing-induced decay modes that scale as
$\epsilon^4$.  These interactions could be discovered through the
observation of displaced decays that are \emph{not} induced by kinetic
mixing --- specifically, a two-fermion decay of spin-0 states (Section
\ref{subsubsec:higgsedDSadditionalCouplings}) and any decay of the
lightest dark-sector fermion (Section \ref{subsec:fermions}).

\subsection{Events from a Higgsed Dark Sector}\label{subsec:higgsedDS}
We assume that the dark sector gauge group $G_D$ has been completely
Higgsed, giving mass to all dark-sector gauge bosons.  We also assume
that all custodial symmetries have been broken, so that couplings
$W_i\, W_j \, W_k$, $W_i\,W_j\,h_k$ and $W_i\,h_i\,h_j$ are generated
between all Higgs and gauge bosons.  The custodial symmetry breaking
ensures that the $A'$ mediates inelastic transitions between
the dark matter and its excited states, which is needed to explain the
DAMA/LIBRA and INTEGRAL signals
\cite{ArkaniHamed:2008qn,Baumgart:2009tn}.  As all mass eigenstates
are mixed, we will not distinguish between the Abelian gauge boson
$A'$ and non-Abelian gauge bosons, but refer to all as $W_D$.

The high-energy collider phenomenology of these models has been
discussed in \cite{Baumgart:2009tn}, and many of the same features
control low-energy collider signatures as well.  The special case of a
pure $U(1)$ dark sector was recently discussed in
\cite{Batell:2009yf}.  In Section \ref{subsubsec:higgsedmeta}, we
determine the metastable states and their decays mediated through
kinetic mixing.  We discuss the final-state multiplicities in Section
\ref{subsubsec:HiggsedEventShapes}.  In Section
\ref{subsubsec:higgsedDSadditionalCouplings}, we comment on other
interactions that induce faster decays than those mediated by kinetic
mixing.

\subsubsection{Metastable States and their Decays through Kinetic Mixing}\label{subsubsec:higgsedmeta}
We begin with a somewhat careful discussion of the metastable states
--- those that can only decay through kinetic mixing to Standard Model
states, or equivalently those whose lifetimes vanish in the $\epsilon
\rightarrow 0$ limit.  States that can decay to final states within
the dark sector are \emph{not} considered metastable, and these decays
are typically prompt.  The Standard Model states produced in the decay
of metastable states include mainly leptons and pions.  We denote the
mass of the lightest dark-sector gauge boson by $\mWm$, and the mass
of the lightest Higgs by $\mhm$.  Any Higgs heavier than $2\mWm$ can
decay to two gauge bosons, and any $W_D$ heavier than $2\mhm$ can
decay to two Higgses.  Therefore, if $\mhm > 2\mWm$, only $W_D$'s are
metastable; if $\mWm > 2\mhm$, then only Higgses are metastable.  More
than one dark gauge or Higgs boson may be metastable in each case, but
they will decay similarly.  If $\f{1}{2} \mWm < \mhm < 2\mWm$, then
the dark sector contains at least one metastable Higgs, and at least
one metastable gauge boson.

\begin{figure}[th]
\begin{center}
\includegraphics[height=2in]{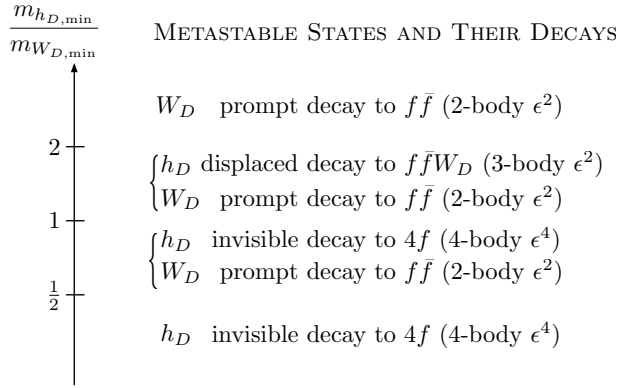}
\caption{
Summary of decay phenomenology in Higgsed dark sectors.
While decays within the dark sector are prompt, the lightest dark-sector state
can only decay into Standard Model leptons or hadrons with a width that
is suppressed by some power of $\epsilon$.
Generically, this decay is prompt if the lightest state is a gauge boson, while it is
displaced or invisible if it is a Higgs.
The figure summarizes the decay phenomenology as a function of the mass ratio of
the lightest Higgs boson to the lightest gauge boson, $\f{\mhm}{\mWm}$.
\label{fig:higgsedDecayStory}}
\end{center}
\end{figure}

The importance of this characterization is that the gauge and Higgs
bosons have very different lifetimes: gauge bosons typically decay
promptly, while Higgs decays give rise to displaced decays or escape
the detector before decaying.  A typical dark-sector production event will
eventually decay to a collection of metastable states, consisting either of
only Higgs bosons, only gauge bosons, or a combination of both.
This leads to a variety of scenarios with very different decay phenomenology,
summarized in Figure \ref{fig:higgsedDecayStory}.
In the following, we justify the decay properties claimed in that figure.

If $W_D$'s are metastable, they can decay to Standard Model
leptons and pions through kinetic mixing.
By assumption, all gauge bosons have some mixing angle $\theta$ with
$U(1)_D$, giving rise to a decay width
\be\label{eq:WDdecay}
\Gamma(W_D\to \ell^+\ell^-) & = & \f{1}{3}\epsilon^2 \theta^2 \alpha m_{W_D} \sqrt{1-\f{4 m_\ell^2}{m_{W_D}^2}} \left(1+\f{2m_\ell^2}{m_{W_D}^2}\right) N_{\rm eff}\nonumber \\
& \simeq & 2.4\;{\rm eV}\;\left(\f{\epsilon}{10^{-3}}\right)^2
\,\theta^2 \,\left(\f{m_{W_D}}{1\,{\rm GeV}}\right)\, \sqrt{1-\f{4 m_\ell^2}{m_{W_D}^2}} \left(1+\f{2m_\ell^2}{m_{W_D}^2}\right) N_{\rm eff},
\ee
where $m_\ell$ denotes the mass of a lepton \emph{or} a hadron.
Here $N_{\rm eff}$ counts the number of available decay products:  
$N_{\rm eff} = 1$ for $m_{W_D} \lesssim 2 m_{\mu}$ when only $W_D\to e^+e^-$ decays 
are possible, and $2+R(m_{W_D})$ for $m_{W_D}\ge2 m_{\mu}$, where 
$R$ is defined to be the energy dependent ratio 
$\sigma(e^+e^- \rightarrow \mbox{ hadrons})/\sigma(e^+e^-  \rightarrow \mu^+\mu^-)$ 
\cite{Amsler:2008zzb}. 

The decay length of $W_D$ in its rest frame is given by
\be\label{eq:WDdecaylength}
c\tau(W_D\to \ell^+\ell^-) \sim 8\times 10^{-6} \; {\rm cm} \;\left(\f{10^{-3}}{\epsilon}\right)^2
\,\theta^{-2} \,\left(\f{1\,{\rm GeV}}{m_{W_D}}\right) \,\left(\f{1}{N_{\rm eff}}\right),
\ee
and is thus prompt for $\theta\sim\OO(1)$.
The combination $\alpha_D \epsilon^2$ can also be normalized to its DAMA/LIBRA
expected value.  In this case,
\be
c\tau(W_D\to \ell^+\ell^-) \sim 10^{-3}\;{\rm cm}\; \theta^{-2} \, \bigg(\f{1\,{\rm GeV}}{m_{W_D}}\bigg) \,
\bigg(\f{1\,{\rm GeV}}{m_{A'}}\bigg)^4 \, \bigg(\f{\alpha_D}{\alpha}\bigg) \,
\bigg(\f{1\,{\rm TeV}}{m_{DM}}\bigg) \,\left(\f{1}{N_{\rm eff}}\right),
\ee
where $m_{DM}$ is the mass of the dark matter (see Section \ref{sec:DAMANormalization}).
We see that prompt decays are also generic.
However, due the strong dependence of $c\tau$ on $1/m_{A'}^{4}$, displaced vertices
are also possible for $m_{A'}\sim\OO$(100 MeV).

\begin{figure}[t]
\begin{center}
\includegraphics[width=3.0in]{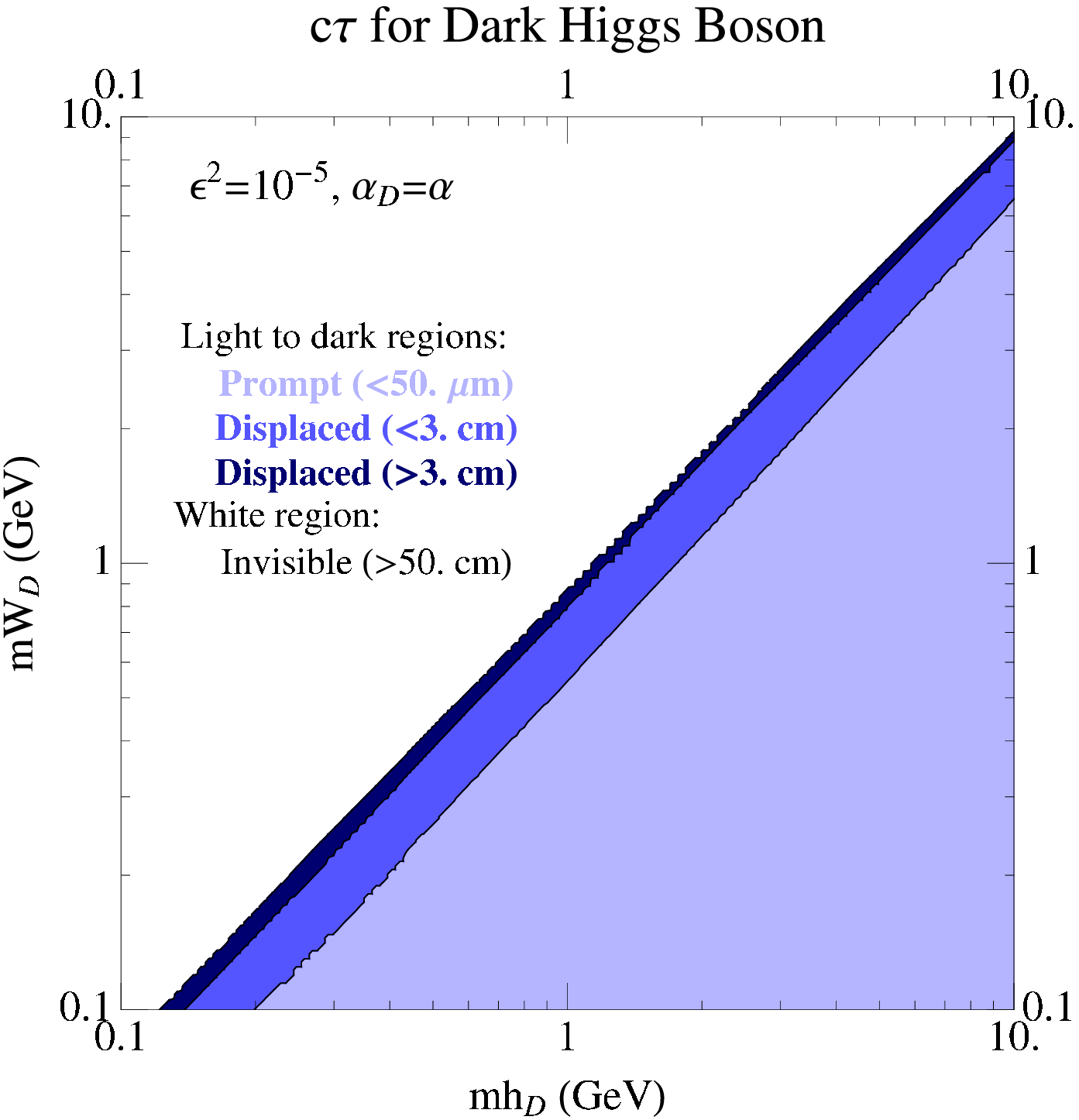}
\includegraphics[width=3.0in]{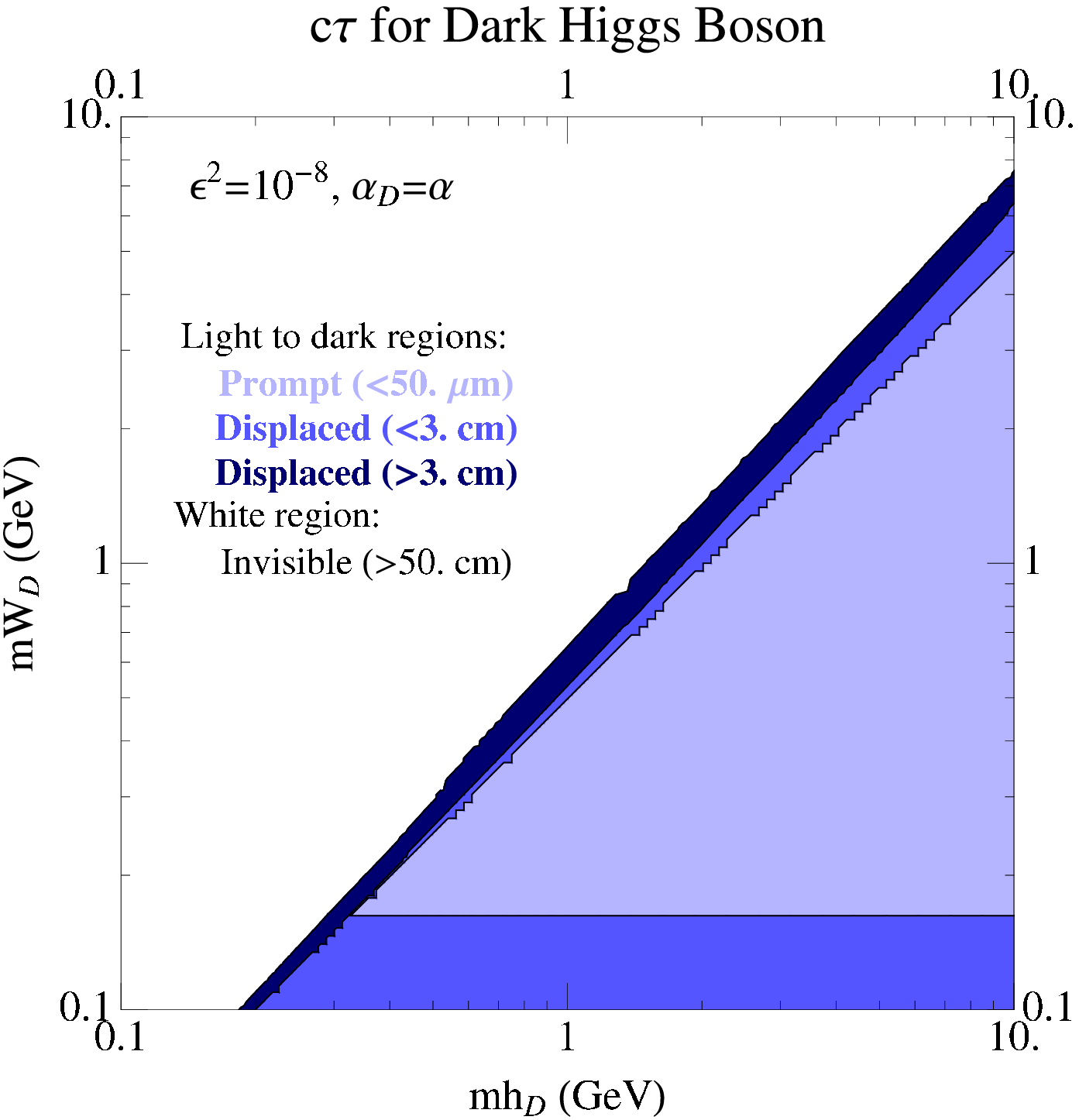}
\caption{Decay length ($c\tau$) of a dark-sector Higgs boson,
$h_D\rightarrow W_i W^*_j\rightarrow W_i \ell \ell$, as a function of
its mass $m_{h_D}$ and the dark-sector gauge boson mass $m_{W_D}$, for $\alpha_D=\alpha$ and
$\epsilon=10^{-5}$ (left) and $\epsilon=10^{-8}$ (right).
Here $\ell$ denotes a Standard Model lepton or hadron.
The light to dark (light-blue/blue/dark-blue) regions indicate decays that are
prompt ($ < 50$ $\mu$m), short displaced ($< 3$ cm), and long displaced ($> 3$ cm), while
in the white region the decays are invisible ($> 50$ cm).
\label{fig:higgsedDecayLifetimes}}
\end{center}
\end{figure}

\begin{figure}[th]
\begin{center}
\includegraphics[width=3.0in]{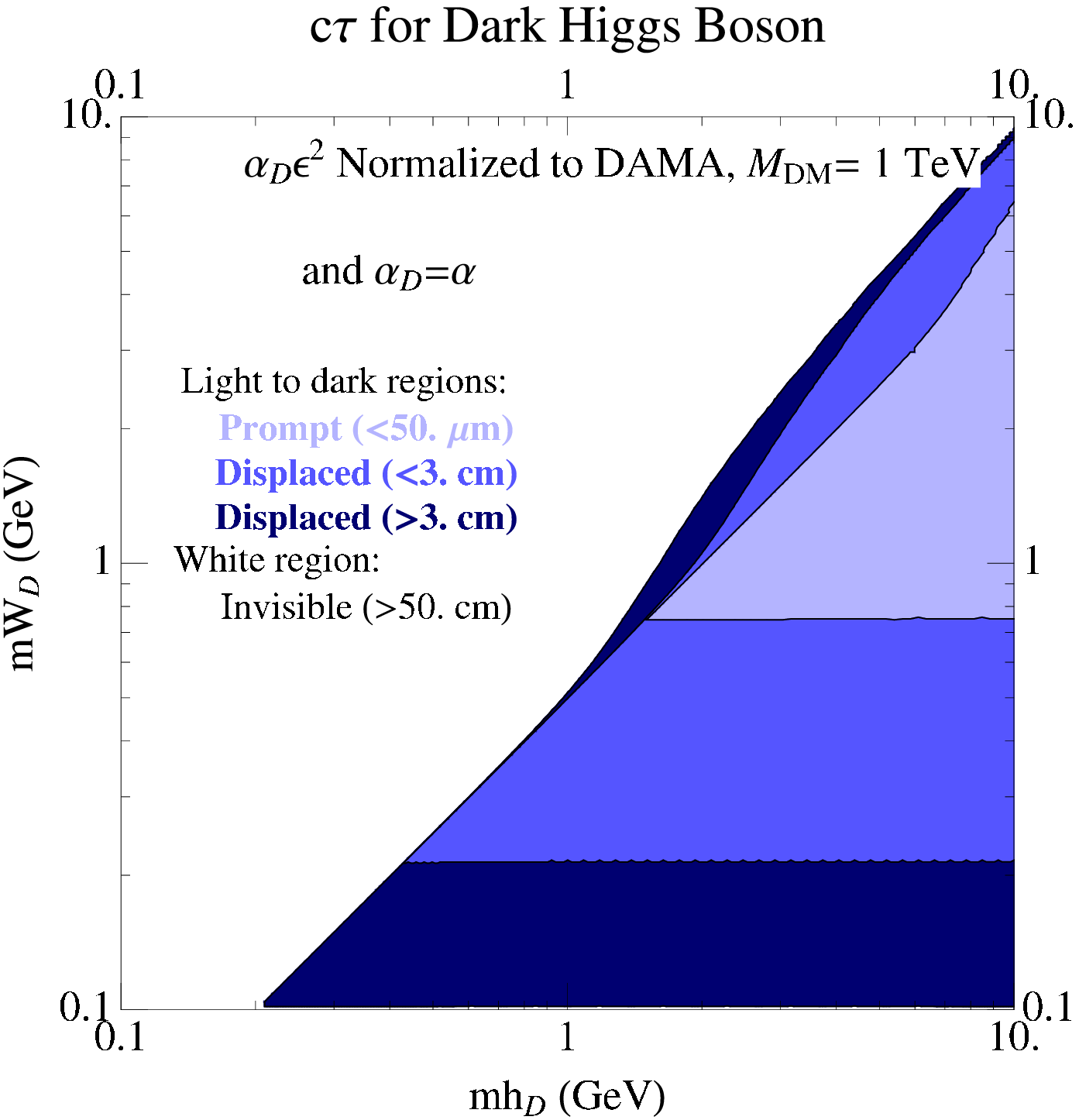}
\caption{Decay length ($c\tau$) of a dark-sector Higgs boson,
$h_D\rightarrow W_i W^*_j\rightarrow W_i \ell \ell$, as a function of
its mass $m_{h_D}$ and the dark-sector gauge boson mass $m_{W_D}$, for $\alpha_D=\alpha$.
Here the combination $\alpha_D\epsilon^2$ has been normalized to its DAMA/LIBRA expected value,
assuming that the dark matter (with mass 1 TeV) is part of the Higgsed dark sector and charged under
the $U(1)_D$ (see Section \ref{sec:DAMANormalization} for further explanation).
The shaded regions in this figure are the same as in Figure \ref{fig:higgsedDecayLifetimes}.
\label{fig:higgsedDecayLifetimesDAMA}}
\end{center}
\end{figure}

If Higgses are metastable, they can decay to Standard Model leptons and hadrons
through one or two off-shell $W_D$'s (heavier Higgses can have 3-body decays to
a lighter Higgs and $W_D^*\rightarrow \ell^+\ell^-$).
If $m_{h_D} > \mWm $, one final-state $W_D$ can be on-shell, and the lifetime is
\be\label{eq:threebody}
\Gamma(h_D\rightarrow W_i W^*_j\rightarrow W_i \ell^+ \ell^-) \simeq
\frac{\alpha \alpha_D \epsilon^2}{128\pi} m_{h_D} f_3(m_{W_i}/m_{h_D}),
\ee
where we have assumed $m_{W_i}=m_{W_j}$ for simplicity.
This has the same form as the $h \rightarrow Z Z^*$ partial width in
the Standard Model.
The function $f_3(x)$, $\frac{1}{2} < x < 1$, is a 3-body phase space integral
that can be found e.g.~on page 30 in \cite{Gunion:1989we}.
The decay length in the rest frame of $h_D$ is shown in
Figure \ref{fig:higgsedDecayLifetimes} for $\alpha_D=\alpha$ and for both
$\epsilon = 10^{-5}$ and $10^{-8}$.
From the figure, we see that $h_D$ decays can be prompt, displaced or invisible,
depending on the parameters.
Similarly diverse decay scenarios are also shown in
Figure \ref{fig:higgsedDecayLifetimesDAMA}, where we plot the $h_D$ decay length
after normalizing $\alpha_D \epsilon^2$ to its DAMA/LIBRA expected value.

If $\mhm < \mWm$, then the lightest Higgs can only decay
through two off-shell gauge bosons, which is dominated by a
four-fermion final state, and is parametrically given by
\be
\Gamma(h_D \to W_D^* W_D^* \to 4\ell ) \sim \frac{\alpha_D \alpha^2 \epsilon^4}{12 \pi}.
\ee
A 2-body decay to two Standard Model fermions through a one-loop diagram mediated
by $W_D$'s is parametrically of the same size.
Because of the strong $\epsilon^4$ suppression, such a light Higgs is expected to
escape the detector, even for large $\alpha_D$.  Note that in this case all
but the very lightest Higgs have three-body decays to either a lighter gauge
boson or a lighter Higgs, so that it is not hard to construct dark
sectors in which prompt, displaced, and invisible decays all occur.


\subsubsection{Event Shapes in a Higgsed Dark Sector}
\label{subsubsec:HiggsedEventShapes}
Specific choices for the structure of a Higgsed dark sector can
dramatically change the typical decay chains within the
sector. It is typical for dark-sector
production to result in a high-multiplicity and lepton-rich final
state.  Here, we discuss two examples.

\begin{figure}[th]
\begin{center}
\includegraphics[width=0.4\textwidth]{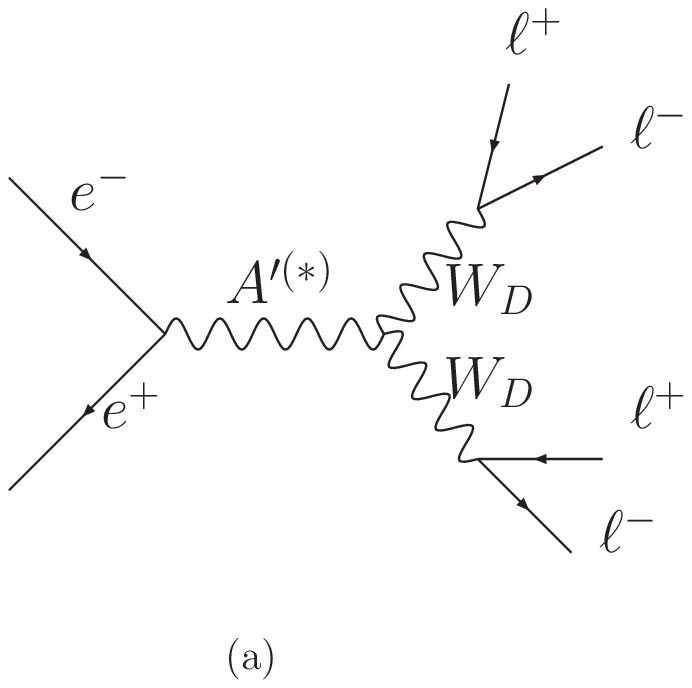}
\hskip 8mm
\includegraphics[width=0.4\textwidth]{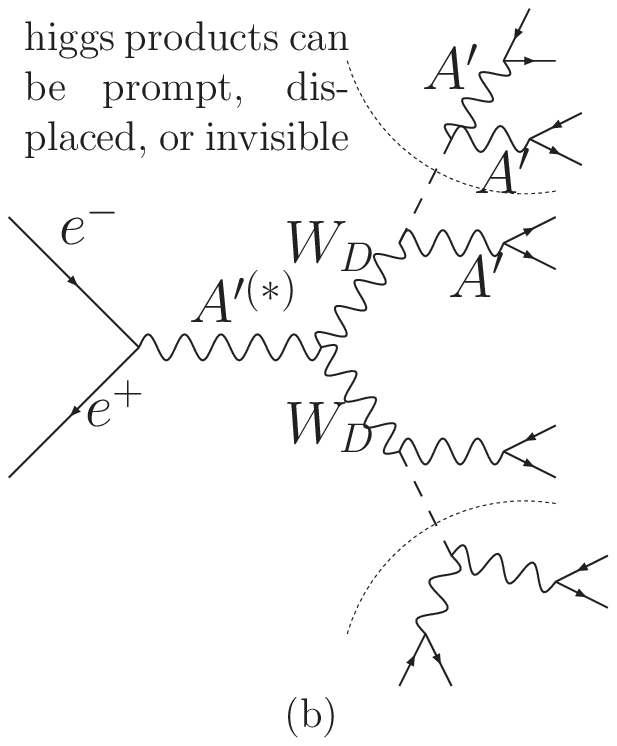}
\caption{\label{fig:feynHiggsed} Two examples of cascade decay chains for Higgsed
  dark sectors.  Left: The production of a pair of dark-sector gauge bosons $W_D$ can
  lead to a four-lepton event.  Right: The production of a pair of heavy dark-sector
  gauge bosons, which decay within the dark-sector to lighter gauge and Higgs bosons, can
  lead to an event with a large number of leptons (and possibly pions).}
\end{center}
\end{figure}

A four-lepton final state is expected from production of a pair of
metastable $W_D$ bosons that decay
directly to Standard Model fermions (Figure \ref{fig:feynHiggsed}(a)).
The lepton kinematics depends on the $W_D$ masses: if $m_{W_D} \ll
\sqrt{s}$, then the observed leptons will lie in two approximately
collinear pairs.

It is also possible to produce heavier gauge bosons that decay into a
lighter $W'_D$ and $h_D$ of the dark sector, as in Figure
\ref{fig:feynHiggsed}(b).
One limit in which this is guaranteed is a dark sector with the
structure of a charge-breaking Standard Model, with $SU(2)_D\times
U(1)_D'$ Higgsed to $U(1)_D$ at $m_{W_D}$ and with charge ($U(1)_D$)
being broken at a lower mass scale $m_{A'}$.  Such models have an
approximate global custodial symmetry $U(1)$ that suppresses the
charged $W_D$'s mixing with the $A'$, so that the heavy $W_D$ bosons
are produced far more efficiently than the lighter $A'$
\cite{Baumgart:2009tn}.  In this case, a large number of leptons (and
possibly pions) are expected.  Likewise, it is possible to produce
dark Higgs bosons directly (a special case is the
``Higgs$'$-strahlung'' discussed in \cite{Batell:2009yf}).

\subsubsection{Decay Modes from Additional Mixing
  Interactions}\label{subsubsec:higgsedDSadditionalCouplings}
Additional couplings between the Standard Model and dark sector are
possible, but they involve massive Standard Model-charged states and so
are irrelevant to dark-sector production.  They can, however, compete
with the $\epsilon^4$-suppressed decays of any light Higgs bosons in
the dark sector.
For example, the mixed Higgs quartic
\be\label{eq:higgsquartic}
\delta\mathcal{L} = \lambda_\epsilon |h_D|^2 |h_{SM}|^2
\ee
leads to Higgs decay through mixing with the
Standard Model Higgs.  If the theory is supersymmetric at high
energies, $\lambda_\epsilon\sim \epsilon g g_D$ is generated by
D-term mixing \cite{Baumgart:2009tn,Katz:2009qq,Cheung:2009qd}.
More generally, loops of link fields charged under $G_D$ and $SU(2)_L\times U(1)_Y$
and loops of ordinary Standard Model matter will generate a quartic Higgs
mixing through their Yukawa interactions with both Higgses.
A rough upper bound on the mixing,
\be\label{eq:naturalness}
\lambda_\epsilon \lesssim \f{m_{h_D}^2}{v_{SM}^2},
\ee
can be obtained by demanding that $m_{h_D}$ receives at most
$\OO$(1) corrections from this quartic term.  If the
$U(1)_D$ symmetry-breaking is generated by (\ref{eq:higgsquartic}),
then this limit is saturated.

When $h_D$ gets a vacuum expectation value, the quartic coupling (\ref{eq:higgsquartic})
generates mixing between $h_D$ and $h_{SM}$ with angle
\be
\theta \sim \f{\lambda_\epsilon}{\lambda_{SM}} \f{v_D}{v_{SM}}
\rightarrow \f{\lambda_\epsilon}{\lambda_{SM}} \sqrt{\f{\lambda_\epsilon}{\lambda_{D}}},
\ee
where the latter limit corresponds to saturation of the bound
\eqref{eq:naturalness}.  This mixing leads to a decay width
\be
\Gamma({h_D\rightarrow \ell^+\ell^-})_{mix} \sim \f{y_\ell^2}{4\pi} m_{h_D} \theta^2
\rightarrow \f{y_\ell^2}{4\pi} m_{h_D} \frac{\lambda_{SM}}{\lambda_D}
\left( \f{m_{h_D}}{m_{h_{SM}}}\right)^6,
\ee
where the final expression is again obtained by saturating
\eqref{eq:naturalness}.
Assuming saturation, the decay length of $h_D$ in its rest frame is given by
\be
c\tau(h_D\rightarrow \ell^+ \ell^-)_{mix} \sim 7 \;{\rm cm}\; \lambda_D\,\bigg(\f{y_\ell}{y_\tau}\bigg)^{-2}\,
\bigg(\f{m_{h_D}}{3\,{\rm GeV}}\bigg)^{-7}\, \bigg(\f{m_{h_{SM}}}{120\,{\rm GeV}}\bigg)^{4},
\ee
where $y_\tau\simeq 0.01$ is the $\tau$ Yukawa coupling.
This decay can thus lead to observable decays with $\OO$(cm) displacements if $h_D$ is
above the $\tau$ threshold, and saturates the naturalness bound.
It dominates over the three-body decay (\ref{eq:threebody}) if $\epsilon \lesssim 10^{-4}$.
Below the $\tau$ threshold, however, it is unlikely to produce observable decays.

\subsection{Events from a Confined Dark Sector}\label{subsec:confinedDS}
In this section, we discuss the events obtained with confined dark
sectors.  We begin in Section \ref{subsec:CDSspectroscopy} with a
review of the spectroscopy in confined dark sectors containing one or
multiple light flavors, and discuss the shape of production events.
As for the Higgsed case, the events can usefully be characterized by
discussing the metastable states, i.e.~those states that do not decay
to other dark-sector particles.  For multiple light flavors, these
metastable states are either the dark-sector pions (Section
\ref{sec:twoFlavor}) or the $A'$ (Section \ref{sec:lightAPrimeCase}),
depending on which is lighter.  For a single flavor, the metastable
states are either the $A'$ (also Section \ref{sec:lightAPrimeCase}),
or the tower of mesons lighter than $2m_{\eta'_D}$, where
$m_{\eta'_D}$ is the mass of the lightest meson (Section
\ref{sec:oneFlavorTower}).  Likewise, the lightest baryon $\Delta_D$
is exactly stable and a tower of baryons lighter than $m_{\Delta_D} +
m_{\eta'_D}$ can be metastable.

\subsubsection{Preliminaries: Spectroscopy and Event Shapes}\label{subsec:CDSspectroscopy}
We begin by reviewing the spectra of mesons and baryons in a
one-flavor model.  A model with only one light flavor has no light
pion; instead, the lightest meson is expected to be a pseudoscalar
meson ($J^{PC} = 0^{-+}$), which we will call $\eta'_D$ in analogy
with the Standard Model $\eta'$.  The second-lightest meson, again
named by analogy with the Standard Model, is the $\omega_D$
($1^{--}$).  Excited states and higher-spin mesons are also expected,
but their spectrum is not known precisely.  We will only ever be
interested in the tower of mesons with mass less than $2 m_{\eta'_D}$.
Mesons above this threshold can decay strongly to lighter mesons, and
are quite broad.  The one-flavor dark sector also contains a tower of
baryons, beginning from the spin-$N_c$/2 $\Delta_D$, where $N_c$ is
the number of colors of the confining gauge group.  For small $N_c$,
these have comparable mass to the tower of light mesons.  We expect
that any metastable glueball states will mix with the mesons, and
decay through this mixing.

In a model with $N_f > 1$ light flavors, confinement breaks an approximate
$SU(N_f)_L \times SU(N_f)_R \times U(1)$ flavor symmetry to a diagonal subgroup
$SU(N_f)_d \times U(1)$.  There are then $N_f^2-1$ light pions.
All other mesons and (for $N_c>2$) all baryons are
parametrically heavier than the pions, and are produced much more
rarely; therefore, we will focus on the pions.

The final-state kinematics and multiplicities can be estimated from
those in the Standard Model by analogy with $`e^+e^- \rightarrow$
hadrons' processes and a simple rescaling of the QCD to
dark-sector confining scale, or inferred from
parametrizations such as the quark production rule \cite{Xie:1988wi}.
These properties depend dominantly on the ratio of the confining scale to
the energy $\sqrt{s_h}$ going into the confined sector ($\sqrt{s_h} =
E_{\rm cm}$ for $e^+e^- \rightarrow \mbox{hadrons}$ or $e^+ e^-
\rightarrow \mbox{dark-sector hadrons}$, while $\sqrt{s_h} = m_{A'}$ for the radiative
return process).  We will parametrize this by $R_E \equiv \sqrt{s_h}/m_{\eta'_D}$.
Thus, for example, from low-energy SPEAR data we see
that jet structure is not observable for $R_E \approx 4$ (3.6 GeV
collisions), but is observable by $R_E \approx 8$
\cite{Schwitters:1975dm}.

Likewise, dark-hadron multiplicities can be estimated from QCD track
multiplicities in $e^+e^-$ collisions; what is measured is the
charged-track multiplicities $\vev{n_{ch}}$, not the total hadron
multiplicity, but studies of the total energy of reconstructed tracks
suggest that approximately half of produced hadrons are
charged \cite{Schwitters:1975dm}.  Doubling observed track
multiplicities from e.g.~\cite{Abe:1996vs,Chliapnikov:1992dk}, we
see that over the range of interest, hadron multiplicities range
between $\vev{N} \approx$ 10 at 7.6 GeV ($R_E\approx 8$) to $\approx 40$
at 90 GeV ($R_E\approx 100$).

At low $R_E$ in a one-flavor model in which $\eta'_D$ is the lightest
state, we expect $\vev{N} \sim \sqrt{s}/m_{\eta'_D}$, so that the
production of non-relativistic hadrons dominates, while the production
of a much lower multiplicity of hadrons is exponentially suppressed
(see e.g. \cite{JeanMarie:1976un} for experimental evidence of this
effect in QCD events with very few pions).  It is difficult to make
more quantitative statements in this region of parameter space based
on either direct extrapolation from QCD or parametrized models.  The
QCD extrapolation would suggest hadron multiplicities larger than
$E_{\rm cm}/m_{\eta'_D}$, which are consistent in theories with light
pions but unphysical if the $\eta'_D$ is the lightest state.
Parametrizations such as the quark production rule \cite{Xie:1988wi}
can in principle be modified for the one-flavor scenario, but they
deviate from experimental data precisely at small $R_E$.

\subsubsection{Dark Sectors with Metastable Pions}\label{sec:twoFlavor}
In a model with two or more flavors and a heavy $A'$, the lightest
dark-sector states are pions.
Since they are then kinematically forbidden to decay into any other dark-sector
states, they are metastable.

\begin{figure}[!t]
\begin{center}
\includegraphics[width=3.0in]{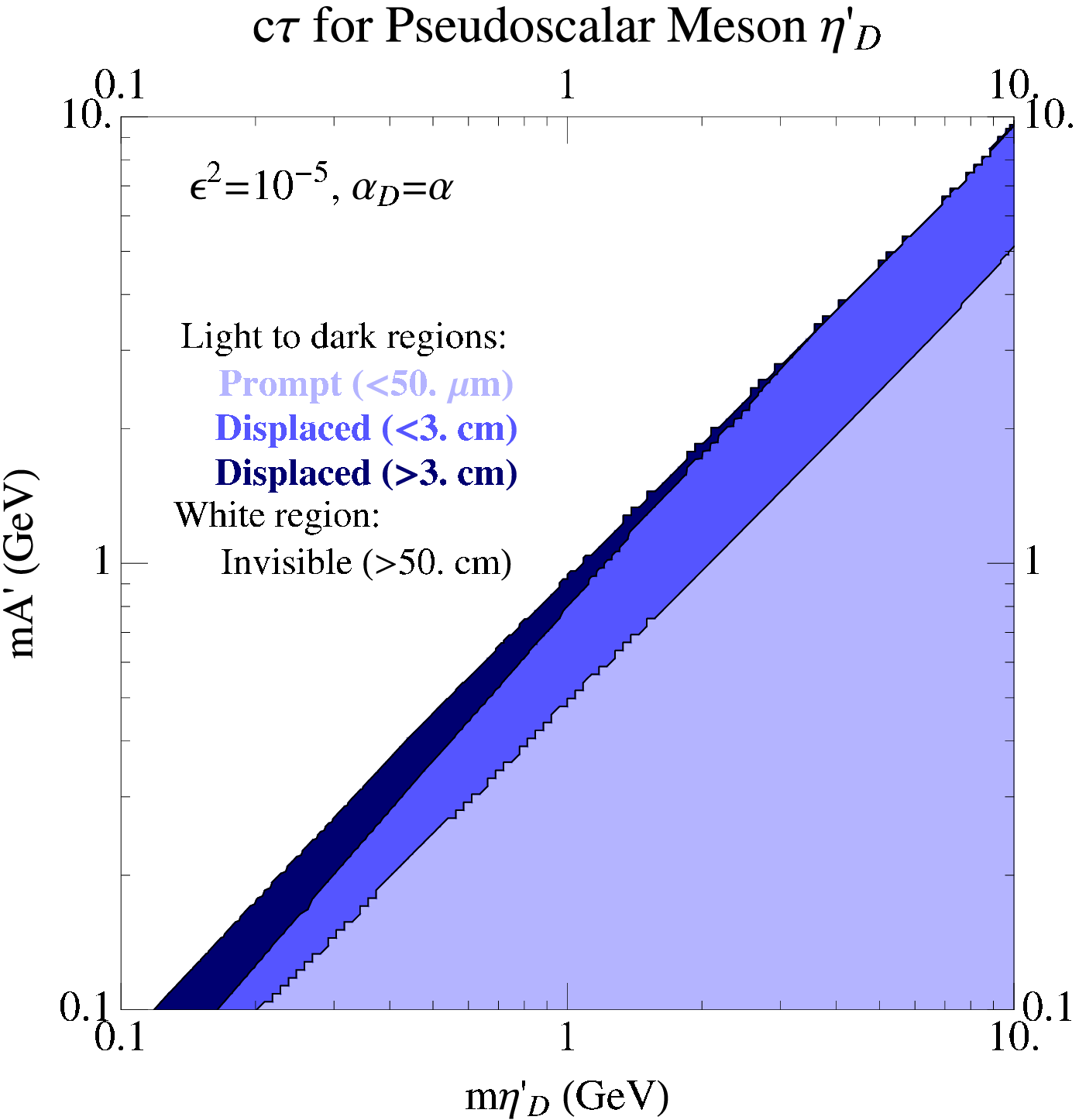}
\includegraphics[width=3.0in]{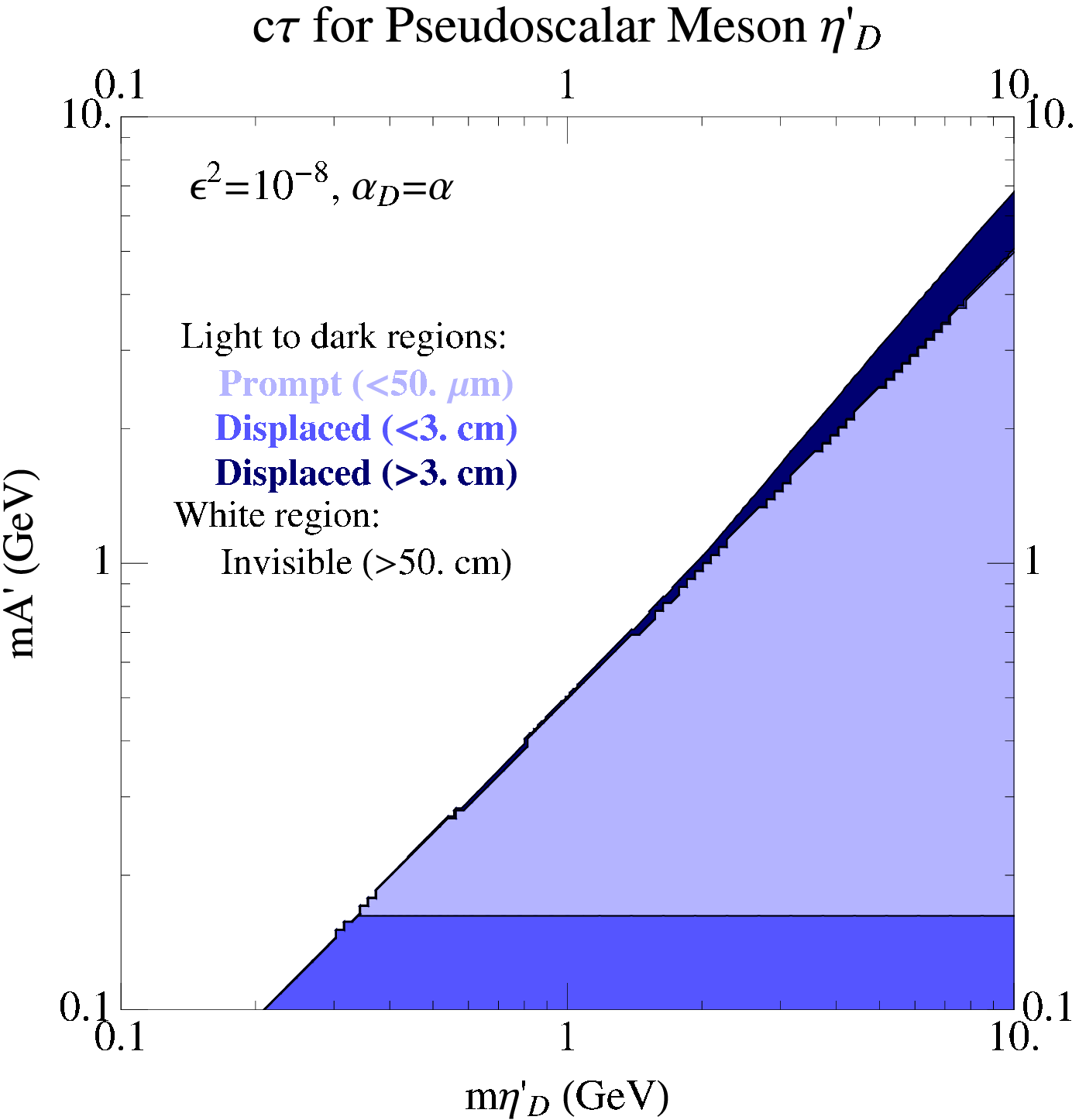}
\end{center}
\caption{\label{fig:etaDecay} Decay length ($c\tau$) of the dark-sector scalar
meson $\eta'_D$, decaying through the axial anomaly to two on- or off-shell
$A'$ gauge bosons, which in turn decay into two Standard Model leptons or
hadrons.  Here $\alpha_D=\alpha$, and $\epsilon=10^{-5}$ (left) or
$\epsilon=10^{-8}$ (right).
The shaded regions in this figure are the same as in Figure \ref{fig:higgsedDecayLifetimes}.
This figure is also applicable to decays of light pions for dark sectors with multiple light flavors.}
\end{figure}

\begin{figure}[!t]
\begin{center}
\includegraphics[width=3.0in]{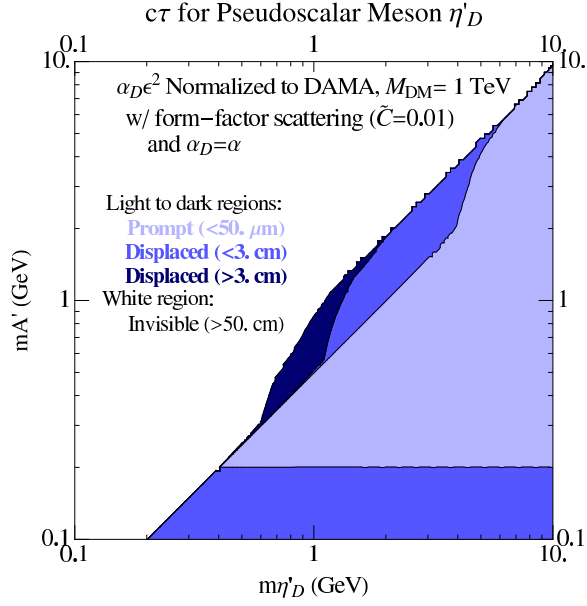}
\end{center}
\caption{\label{fig:etaDecayB} Decay length ($c\tau$) of the dark-sector scalar
meson $\eta'_D$, decaying through the axial anomaly to two on- or off-shell
$A'$ gauge bosons, which in turn decay into two Standard Model leptons or
hadrons.  Here $\alpha_D=\alpha$, and the combination
$\alpha_D\epsilon^2$ has been normalized to its DAMA/LIBRA expected value,
assuming that the dark matter is a 1 TeV dark meson neutral under $U(1)_D$, but
consists of a light and a heavy quark charged under $U(1)_D$
(see Section \ref{sec:DAMANormalization}).
The shaded regions in this figure are the same as in Figure
\ref{fig:higgsedDecayLifetimes}. This figure is also applicable to decays of light pions for dark sectors with multiple light flavors.}
\end{figure}

\begin{figure}[t!]
\begin{center}
\includegraphics[width=3.0in]{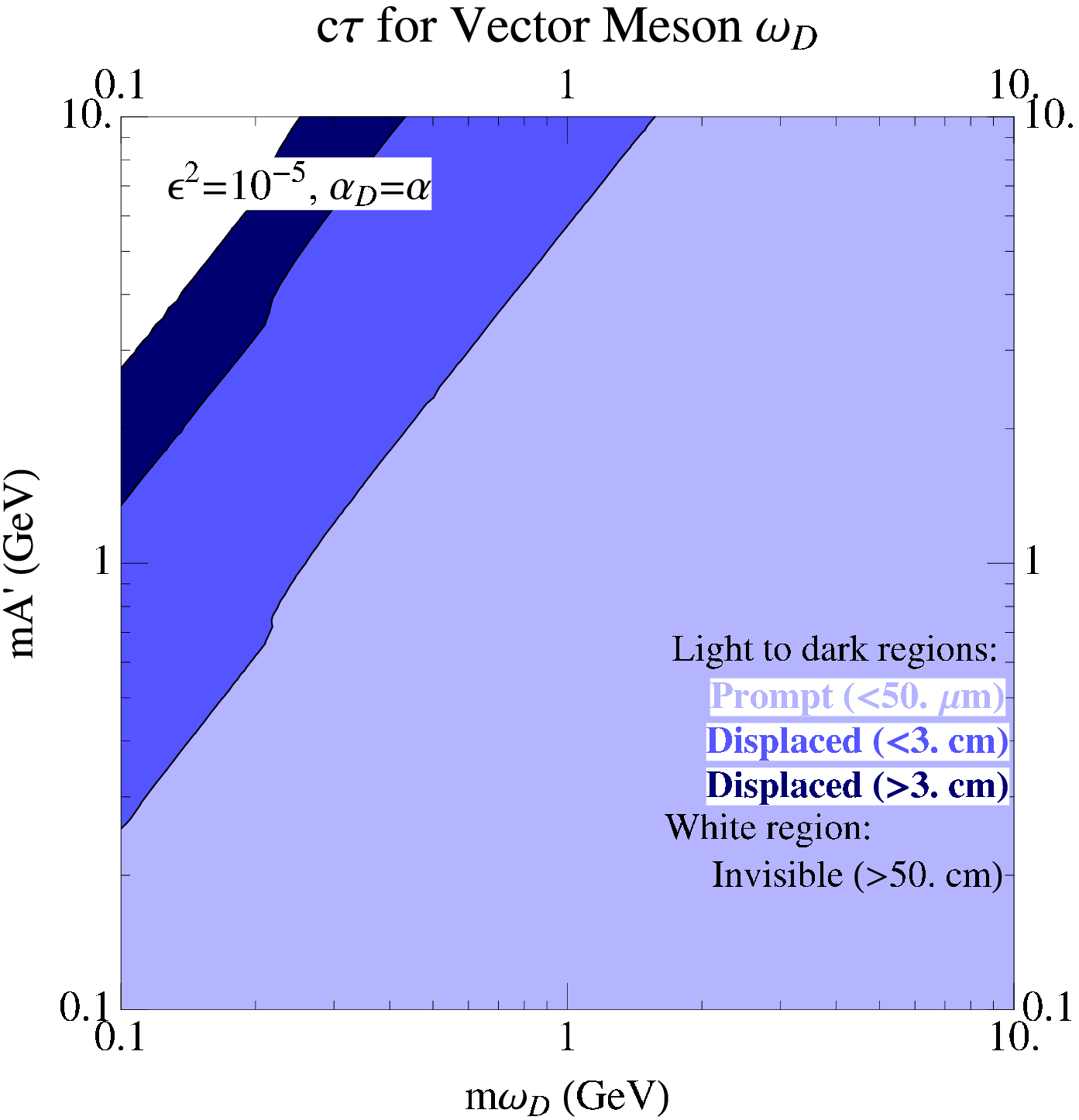}
\includegraphics[width=3.0in]{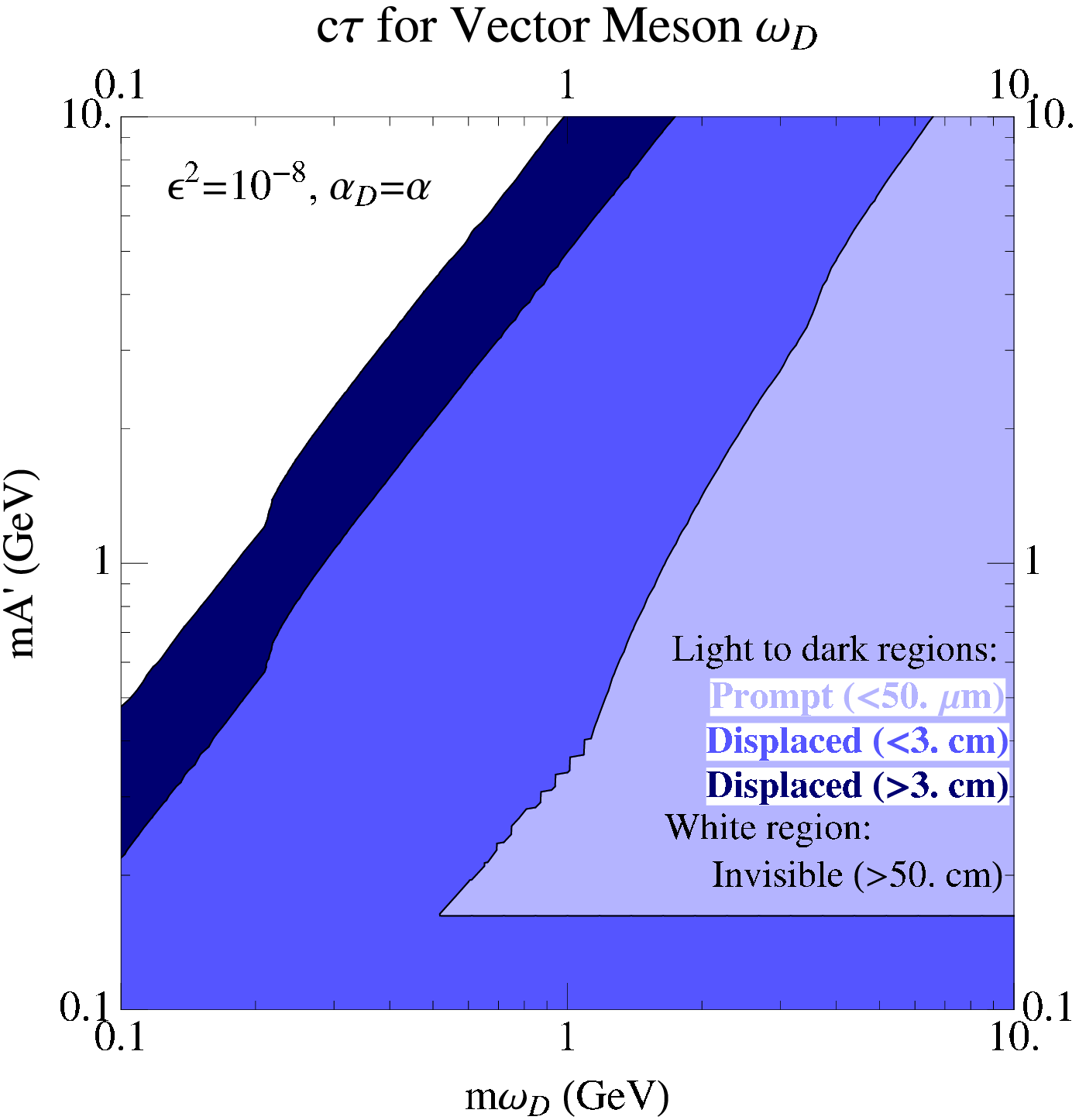}
\end{center}
\caption{\label{fig:phiDecay} Decay length ($c\tau$) of the dark-sector vector
meson $\omega_D$, which mixes with the $A'$ and decays to two Standard Model leptons or
hadrons.  Here $\alpha_D=\alpha$, and $\epsilon=10^{-5}$ (left) or
$\epsilon=10^{-8}$ (right).
The shaded regions in this figure are the same as in Figure \ref{fig:higgsedDecayLifetimes}.}
\end{figure}

\begin{figure}[t!]
\begin{center}
\includegraphics[width=3.0in]{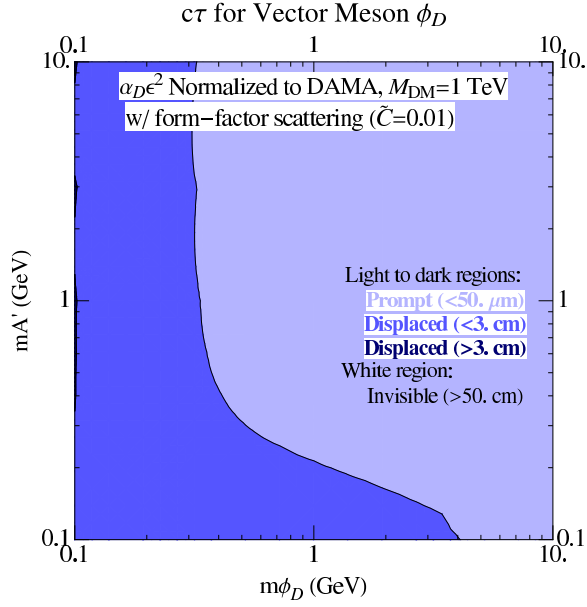}
\end{center}
\caption{\label{fig:phiDecayDAMA} Decay length ($c\tau$) of the dark-sector vector
meson $\omega_D$, which mixes with the $A'$ and decays to two Standard Model leptons or
hadrons.
Here $\alpha_D=\alpha$, and the combination $\alpha_D\epsilon^2$ has been normalized
to its DAMA/LIBRA expected value, assuming that the dark matter is a 1 TeV dark meson
neutral under $U(1)_D$, but consists of a light and a heavy quark charged under $U(1)_D$
(see Section \ref{sec:DAMANormalization}).
The shaded regions in this figure are the same as in Figure \ref{fig:higgsedDecayLifetimes}.}
\end{figure}

\begin{figure}[t!]
\begin{center}
\qquad\includegraphics[width=0.8\textwidth]{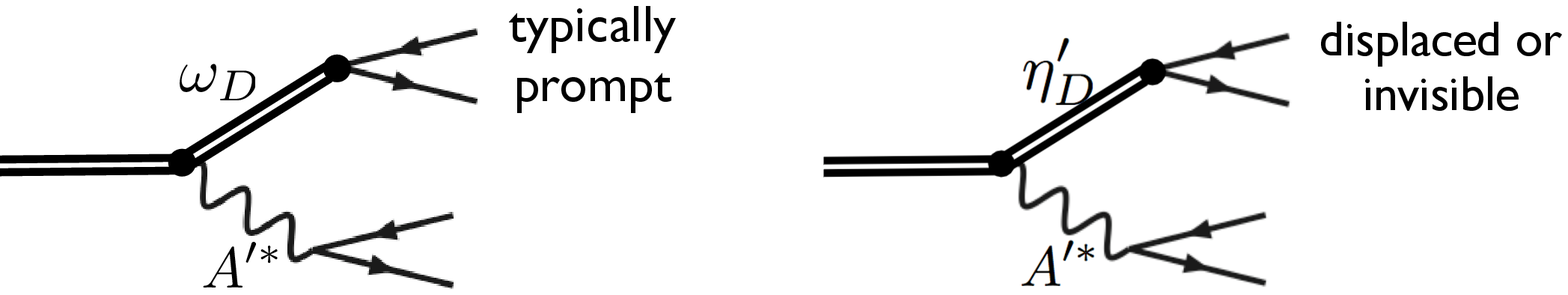}
\vskip 1cm
\includegraphics[width=3.0in]{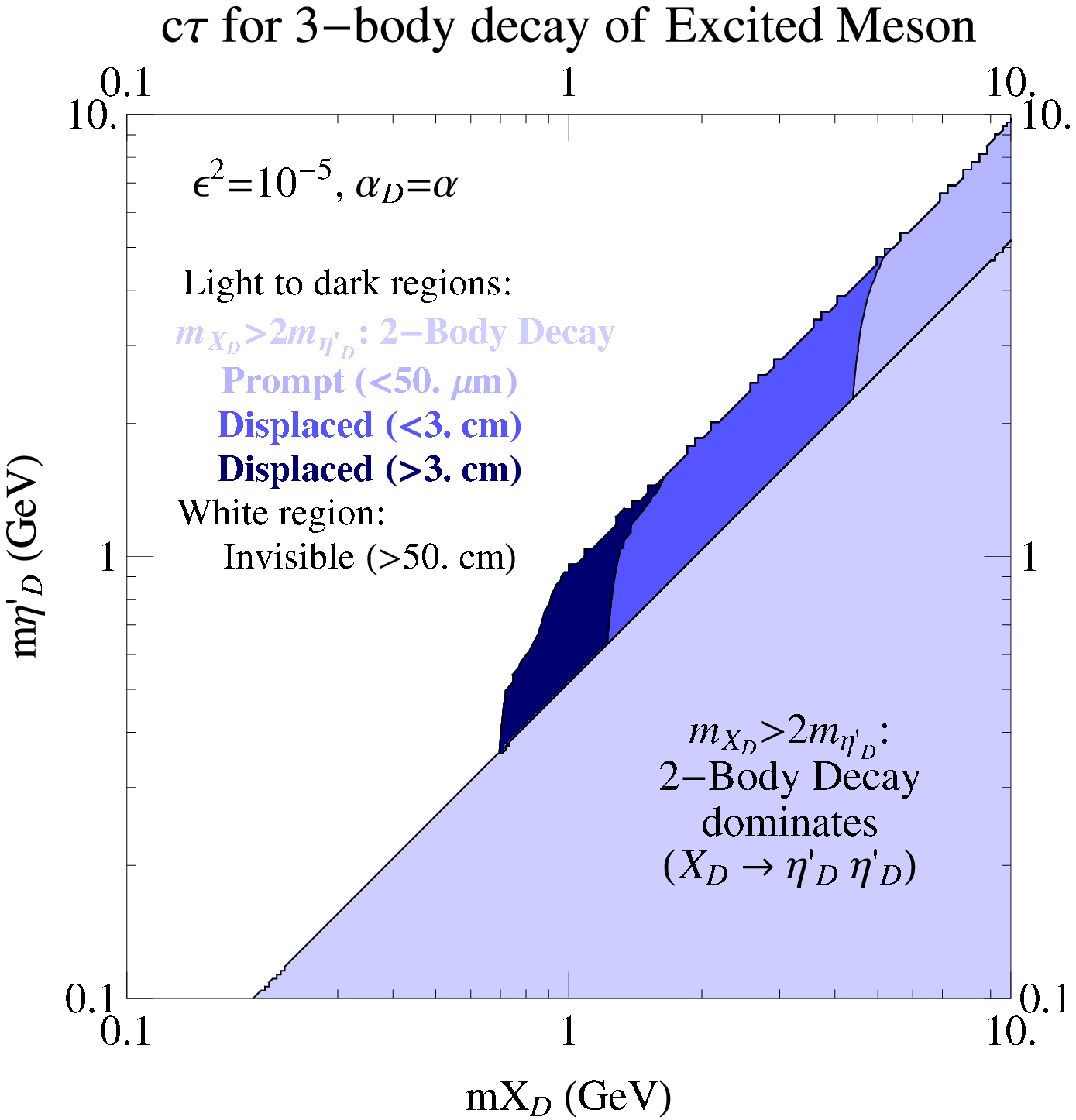}
\includegraphics[width=3.0in]{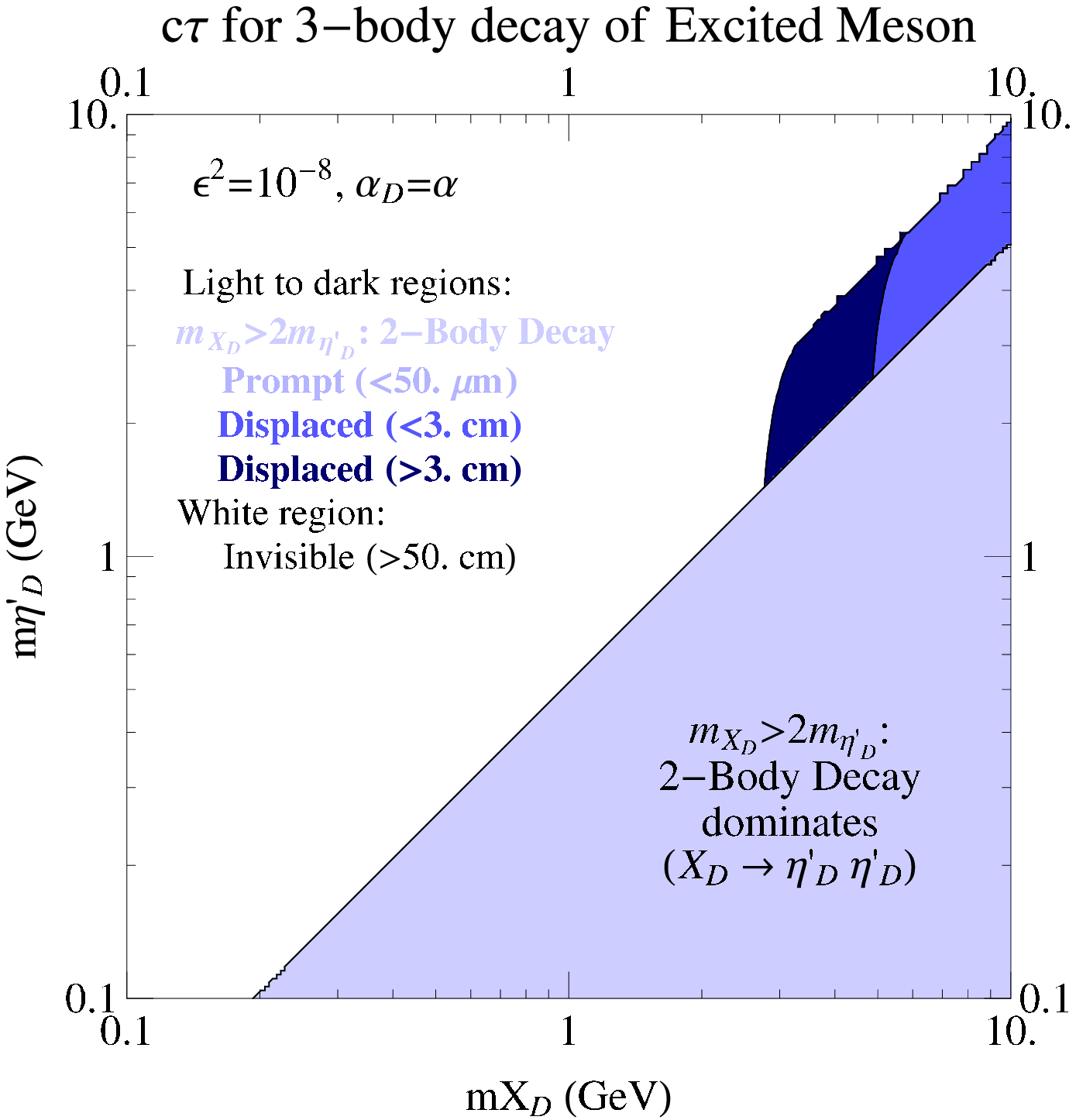}
\end{center}
\caption{\label{fig:threebodyDecay}
Top: Diagram showing 3-body decays of a heavy dark-sector excited spin-0 and spin-2
meson $X_D$ into Standard Model states (through an off-shell $A'$) and a lighter
meson.
Bottom: Decay length ($c\tau$) of $X_D$.
Here $\alpha_D=\alpha$, and $\epsilon=10^{-5}$ (left) or $\epsilon=10^{-8}$ (right).
In the very light blue region (lower right triangle) the 2-body decay of $X_D$ dominates.
The remaining shaded regions in this figure are the same as in Figure \ref{fig:higgsedDecayLifetimes}.}
\end{figure}

\begin{figure}[t!]
\begin{center}
\includegraphics[width=3.0in]{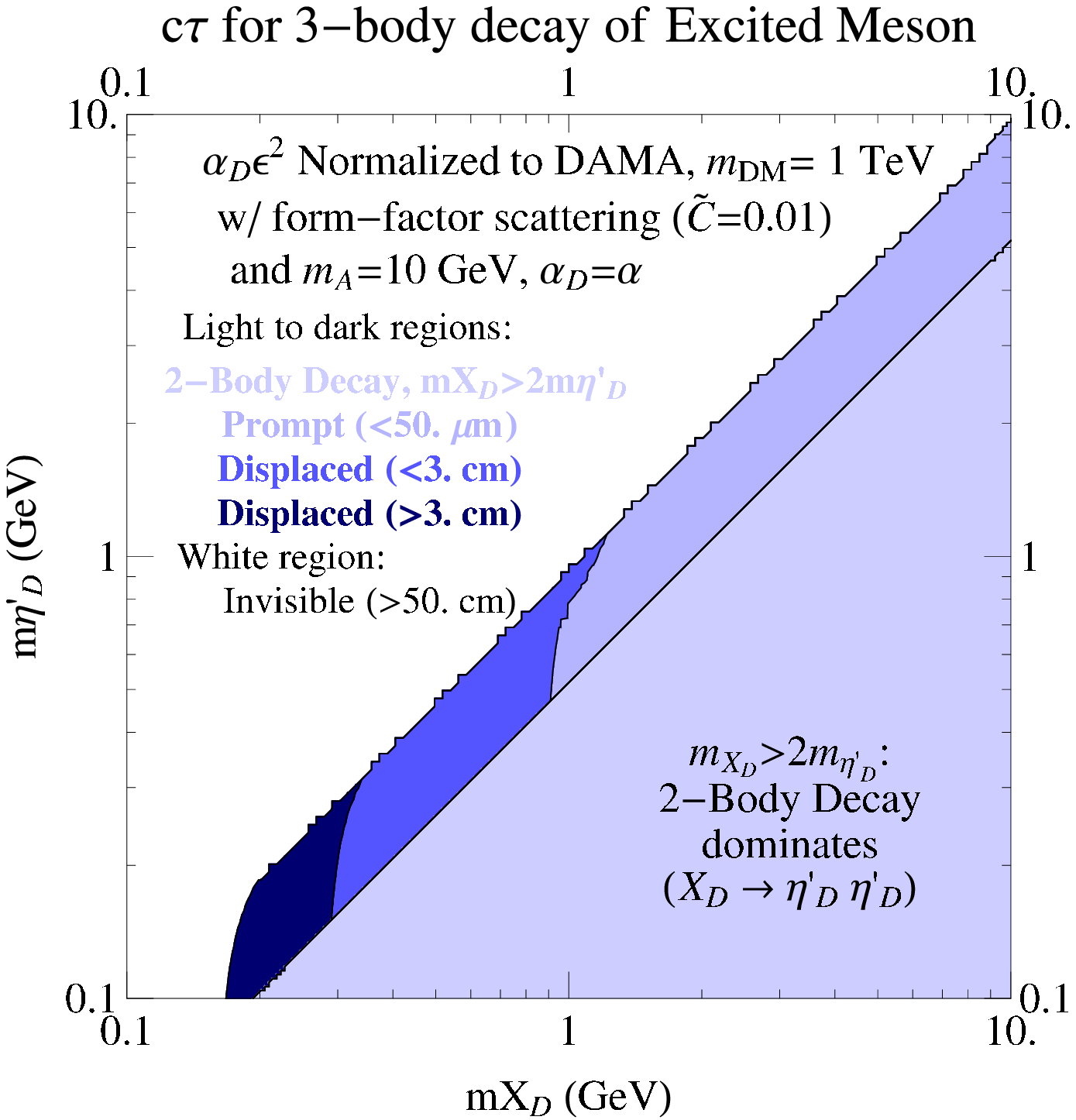}
\end{center}
\caption{\label{fig:threebodyDecayDAMA}
Decay length ($c\tau$) of a heavy dark-sector spin-0 and spin-2 excited meson $X_D$,
which has a 3-body decay to a lighter meson and to two Standard Model
leptons or hadrons through an off-shell $A'$.
Here $\alpha_D=\alpha$, and the combination $\alpha_D\epsilon^2$ has been normalized
to its DAMA/LIBRA expected value, assuming that the dark matter is a 1 TeV dark meson
neutral under $U(1)_D$, but consists of a light and a heavy quark charged under $U(1)_D$
(see Section \ref{sec:DAMANormalization}).
In the very light blue region (lower right triangle) the 2-body decay of $X_D$ dominates.
The remaining shaded regions in this figure are the same as in Figure \ref{fig:higgsedDecayLifetimes}.}
\end{figure}

Kinetic mixing does \emph{not}
generate the most general decays allowed for the dark pion in an
effective field theory.
In particular, the helicity-suppressed decay of a ``$v$-pion'',
$\pi_v \rightarrow \ell^+ \ell^-$, which dominates in Hidden Valley
models with a $Z'$ that couples directly to \emph{both} dark-sector
and Standard Model matter, is not generated by kinetic mixing.
This can be most readily seen before performing the field redefinitions
that diagonalize the gauge field kinetic terms.
Instead, let us treat the kinetic mixing as an insertion
\be
\includegraphics[width=1.2in]{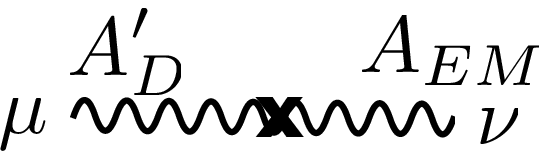} \; \sim \;
(g^{\mu\nu} p^2 -p^\mu p^\nu).
\ee
By Lorentz invariance, the pion current must have the form
\be
\vev{ 0 | A'_{\mu} | \pi(p) } \propto i p_\mu,
\ee
which vanishes when dotted into the kinetic mixing tensor.   In other words,
kinetic mixing only mixes the transverse modes of the two gauge
bosons, to which the pion does not couple.  The vanishing of this
decay mode is obscured when the couplings are diagonalized.  The $A'$
couples to the photon current, $J_{EM}$, but this vector-like current
vanishes when contracted with $p_\mu$.  The $A'$ also couples to the
chiral current, $J_Z$, but this contribution is precisely canceled by
the $Z$'s coupling to the dark photon current, $J_{A'}$, so this also
does not mediate decays.

Instead, the pion can only decay through the anomaly diagram, but with
one or both $A'$ off-shell.  The resulting decay width scales as
\be
\Gamma_{\rm{anomaly}}(\pi_D \to A'^*A'^{(*)}) \sim
\f{\alpha_D^2}{64\pi^3 f_{\pi_D}^2}\,m_{\pi_D}^3\,
\left(\f{1}{4\pi}\alpha\epsilon^2 \f{m_{\pi_D}^4}{m_{A'}^4}
\right)^p
,
\ee
where $p=1$ for three-body decays and $p=2$ for four-body decays, and
the pion decay constant $f_{\pi_D}$ can be estimated by scaling the
pion decay constant in the standard model, $f_{\pi_D}=f_\pi
m_{\eta'_D}/m_{\eta'_{SM}}$.  These decay lifetimes are identical to
those of the $\eta'_D$ in a one-flavor model, and the rest frame
displacements are shown in Figure \ref{fig:etaDecay}.  If $m_{A'} <
m_{\pi_D} < 2 m_{A'}$, a three-body decay can give rise to displaced
vertices, but this mass condition is not generic.  When
$m_{A'}>m_{\pi_D}$, the four-body mode is suppressed by $\epsilon^4$,
leading to very large displacements:
\be
c\tau (\pi_D \to 4\ell) \sim 
\left(\f{\alpha}{\alpha_D}\right)^2\,
\left(\frac{10^{-3}}{\epsilon}\right)^4\,
\left(\frac{1\mbox{ GeV}}{m_{\pi_D}^3/f_{\pi_D}^2}\right)
 10^{12} \mbox{ cm}
\ee  
A loop decay to two leptons is also possible, but comparable in rate
to the four-body decay (in fact, it receives chiral suppression $\sim
(m_{\ell}/m_{\pi_D})^2$ relative to the four-body decay).  Therefore,
in this region of parameter space the pion escapes the detector
without decaying, and is invisible.

Assuming that dark-sector interactions break the dark flavor symmetries,
a heavier pion that cannot decay to two lighter pions may still
decay via an off-shell $A'$ into a lighter pion
and Standard Model leptons/hadrons.
This decay is suppressed by $\alpha_D\epsilon^2$, and so is typically
prompt or displaced.

\subsubsection{Dark Sectors with a Metastable Light $A'$}\label{sec:lightAPrimeCase}

In this section, we discuss the case when the $A'$ is the lightest state
in the dark sector.  We consider both a one-flavor and a multi-flavor
model.  In both cases, final states consisting of prompt Standard Model
leptons and hadrons are likely.

A particularly simple example is the one-flavor model, in which the $A'$
is significantly lighter than the dark mesons and the physical dark Higgs boson
associated with $U(1)_D$ breaking, $h_D$.
Under this assumption, the $\ell^+\ell^- \gamma$ production is important,
but off-shell production of dark-sector particles gives rise to high-multiplicity
final states consisting of leptons and/or Standard Model hadrons that may yield
more sensitive limits.

For light $A'$, all dark mesons and $h_D$ can decay promptly to several $A'$'s.
For example, the $\eta_D'$ decays through the chiral anomaly to $2A'$.
We show the $\eta'_D$ decay length in its rest-frame in Figure \ref{fig:etaDecay}
for two different values of $\epsilon$, and in Figure \ref{fig:etaDecayB} for the
case that $\alpha_D\epsilon^2$ is normalized to its DAMA/LIBRA expected value.
Heavier dark mesons may decay to two or more $A'$'s, or
to an $A'$ and a lighter dark meson.  Baryon number is an accidental
low-energy symmetry of the dark sector, so we will assume that the
$\Delta_D$ is exactly stable, and all other dark baryons decay to it and
one or more $A'$'s.

Therefore, all dark-sector final states eventually decay to a
collection of $A'$'s and stable $\Delta_D$-baryons.
The $A'$ usually decays promptly through kinetic mixing (cf.~equations (\ref{eq:WDdecay})
and (\ref{eq:WDdecaylength})):
\begin{eqnarray}\label{eq:Aprimedecay}
\Gamma(A'\to \ell^+\ell^-) & = & \f{1}{3}\epsilon^2 \alpha m_{A'} \sqrt{1-\f{4 m_\ell^2}{m_{A'}^2}} \left(1+\f{2m_\ell^2}{m_{A'}^2}\right) N_{\rm eff} \nonumber \\
\Rightarrow \;c\tau(A'\to \ell^+\ell^-) & \sim &
8\times 10^{-6} \; {\rm cm} \;\left(\f{10^{-3}}{\epsilon}\right)^2
\,\left(\f{1\,{\rm GeV}}{m_{A'}}\right)\,\left(\f{1}{N_{\rm eff}}\right).
\end{eqnarray}
We expect to produce final states with at least one to two $A'$ per dark hadron.
These decays can easily generate high lepton multiplicities, even when relatively
few dark hadrons are produced.

In one-flavor models at high multiplicities, production
of dark baryons is unavoidable and leads to invisible products which escape
the detector.  For low-multiplicity final states, when $\sqrt{s}/
m_{\eta'} \sim 2$, the efficiency for producing individual
exclusive final states, such as those that do not contain any baryons,
are reasonably high.

Similar comments apply for multi-flavor models, in
which the pions are significantly heavier than the $A'$.  In this
case, some pions may be stabilized by unbroken dark-sector flavor symmetries, or by
an unbroken discrete subgroup of $U(1)_D$ if their charges are non-integer
multiples of the dark Higgs charge.  Kinetic mixing does not change
the stability of these pions, and they will escape the detector.  Pions
that are not stabilized, however, can decay to $2 A'$ through the
chiral anomaly, and these $A'$'s are observable through their prompt
decays to Standard Model matter.

\subsubsection{Dark Sectors with Towers of Metastable Mesons and Baryons}\label{sec:oneFlavorTower}
We now consider the one-flavor model with a heavy $A'$.
If the $A'$ is heavy, then the tower of mesons lighter
than $2 m_{\eta'_D}$ becomes metastable, i.e.~they have no
available decays within the dark sector.
These decays are therefore suppressed by at least $\epsilon^2$,
and can often be displaced or even invisible.
Likewise, all baryons lighter than $m_{\Delta_D} + m_{\eta'_D}$
are metastable for heavy $A'$.

The discussion of dark pion decays in Section \ref{sec:twoFlavor} applies to
the $\eta'_D$ as well.  In particular, decays must proceed through
\emph{two} $A'$ gauge bosons, and if both are off-shell they are
suppressed by $\epsilon^4$.  Therefore, they can easily be invisible.
The decay of $\eta'_D$ through one or two off-shell $A'$'s is included in
Figures \ref{fig:etaDecay} and \ref{fig:etaDecayB}.

The decays of the heavier mesons are generically much easier to see.
The $\omega_D$ vector meson generically mixes with the $A'$, with a
mixing angle that can be estimated from their masses,
\be
\theta \sim \f{m_{\omega_D}^4}{(m_{\omega_D}^2 +
  m_{A'}^2)^2}.
\ee
This leads to a decay width that is suppressed by  $\epsilon^2$, and
up to an unknown $\OO(1)$ coefficient is given by
\be
  \Gamma(\omega_D \rightarrow \ell^+\ell^-) \sim \f{4\pi}{3}\epsilon^2
  \alpha \alpha_D \f{m_{\omega_D}^5}{(m_{\omega_D}^2 + m_{A'}^2)^2}
  \sqrt{1-\f{4 m_\ell^2}{m_{\omega_D}^2}}
  \Big(1+\f{2m_\ell^2}{m_{\omega_D}^2}\Big).
\label{eq:phiDecay}
\ee
The total $\omega_D$ width depends only mildly on $m_{\omega_D}$.
In Figure \ref{fig:phiDecay}, we show $c\tau$ for $\omega_D$ for two different
values of $\epsilon$, while in Figure \ref{fig:phiDecayDAMA}, we
normalize $\alpha_D\epsilon^2$ to its DAMA/LIBRA expected value.

Exotic and excited mesons can have more spectacular decays. Spin-1
mesons are expected to decay into Standard Model pairs, just like the
$\omega_D$, while spin-0 and spin-2 mesons can have 3-body decays into
Standard Model states (through an off-shell $A'$) and a lighter meson,
as shown in Figure \ref{fig:threebodyDecay}.  These have a typical
lifetime, estimated by dimensional analysis, given by
\be
\Gamma(X_D \rightarrow A'^* + (\omega_D/\eta'_D) \to \ell^+\ell^- + (\omega_D/\eta'_D)) \sim
\frac{\alpha_D \alpha \epsilon^2}{12 \pi} \f{m_{X_D}^5}{m_{A'}^4} \sqrt{1- \frac{m_{\omega_D/\eta'_D}^2}{m_{X_D}^2}},
\ee
which can easily be displaced.
This is shown in Figure \ref{fig:threebodyDecay} for two different values
of $\epsilon$, and in Figure \ref{fig:threebodyDecayDAMA} for a case in which
$\alpha_D\epsilon^2$ has been normalized to its DAMA/LIBRA expected value.

Note that, unlike the decay of the $\eta'_D$ into Standard Model states,
where the 3-body region of phase space required a tuning of independent
masses $m_{\eta'_D}$ and $m_{A'}$, the condition $m_{\eta'_D/\omega_D} < m_{X_D}<2m_{\eta_D'}$
is satisfied naturally for a tower of exotic mesons $X_D$.
Similar decays are expected for excited baryons.

To summarize: although significant missing energy from both
baryons and $\eta'_D$'s is generically expected, typical events in
such a sector \emph{also} have a combination of prompt and displaced
decays.

\subsection{Dark Sectors with Uncolored Fermions}\label{subsec:fermions}
We have focused on Higgsed and confined sectors with no light
fermions.  However, the presence of a light fermion changes the
phenomenology of either a Higgsed or a confined dark sector
dramatically.  Some or all of the bosons that are produced can decay
to two dark-sector fermions, instead of decaying to Standard Model
fermions through kinetic mixing.  The dark-sector fermions are
singlets under any confining gauge group (in the confined case, the
dark-sector gauge group may have a subgroup which does not confine).
The lightest dark-sector fermion does not have any decay modes induced
by kinetic mixing, but can decay through new mixing operators of the
form
\be
\f{c}{M^p} (L_D h_D^p) (L h_{SM}),
\ee
where $h_D^p$ is a collection of dark-sector Higgses with the same
quantum numbers as a dark-sector fermion $L_D$, $L$ is a Standard Model fermion,
and $h_{SM}$ is a Standard Model Higgs.
The decay lifetime of the dark fermions is quite
model-dependent, but they will typically be long-lived.  In this case,
any dark-sector bosons that can decay to a pair of light
dark-sector fermions may well be invisible.  The only way to
find such invisible sectors is by observing their recoil against
other particles in the final state (in the context of B-factories, a
photon).

In dark sectors with more than one light fermion, this situation is
ameliorated.  The second-lightest fermion $f_2$ may have decays,
through either Higgs quartic or gauge boson kinetic mixing,
\be
f_2 \rightarrow f_1 A' \to f_1 \ell^+ \ell^-,
\ee
where the $A'$ may be on- or off-shell.  This decay is especially
important when the decay $f_2\to 3f_1$ is kinematically not allowed.
These fermions generically have longer decay lifetimes than their
bosonic counterparts by the ratio of three-body to two-body phase
space, $\sim 1/16\pi^2$.  Such decays are much harder to find than the
bosonic analogues because they contain missing energy, and the
$\ell^+\ell^-$ invariant mass need not reconstruct a resonance.
Instead, the dilepton invariant mass distribution may have an
endpoint.  Moreover, if a sector contains both metastable fermions
\emph{and} metastable bosons, the bosonic decay products may
reconstruct resonances.

\section{Searches at low-energy $e^+e^-$ colliders}\label{sec:searches}
Experiments at high-intensity $e^+e^-$ colliders at low energies, such
as the B-factories BaBar and BELLE, can open a direct window into the
physics of a low-mass dark sector by observing production of
dark-sector particles and their decays back into the Standard Model.
Though striking, these final states are quite different from those of
interest for precision measurements of the Standard Model, and could
have gone unnoticed in existing B-factory datasets.  A common feature
of the dark-sector production processes is a high multiplicity of
leptons, pairs of which reconstruct resonances.  However, the precise
multiplicity and energy of these leptons, and the presence of
displaced vertices or missing energy, depend on the detailed structure
of the dark sector.

This variety makes inclusive searches desirable, but the primary
concern is maintaining acceptance for a large fraction of signal
events while removing Standard Model backgrounds.  To this end, we
identify several search strategies, with exclusive search requirements
for events with large physics backgrounds and inclusive treatments for
more striking signal events.  Our discussion is mostly qualitative; a
quantitative study is beyond the scope of this paper.  Our focus is on
BaBar and BELLE, which represent the current high-intensity frontier,
but similar analyses with data from KLOE, 
CLEO-c, the upcoming BES-III \cite{Asner:2008nq}, and elsewhere could be more sensitive,
depending on the nature of the dark sector.

We identify six regions of interest, where different search
strategies are required:
\begin{itemize}
\item $4\ell$ (exclusive),
  reconstructing $E_{\rm cm}$ (also $4\ell+\gamma$)
\item $4\ell$ (exclusive), with
  displaced dilepton vertices (also $4\ell+\gamma$)
\item $\ge 5\ell + tracks$ (inclusive), reconstructing
  $E_{\rm cm}$ (also + $\gamma$)
\item $\ge 5\ell + tracks$ (inclusive), with displaced
  dilepton vertices (also + $\gamma$)
\item Very high track multiplicity, with many tracks consistent with
  leptons
\item $\gamma$ + nothing
\end{itemize}

\noindent
Before discussing each of these searches in turn, we list three prominent physics backgrounds
that contribute to the channels with four or more leptons:
\begin{enumerate}
\item QED processes, $e^+e^- \rightarrow 4\ell$, which can reconstruct the full
  beam energy.
\item Two photons, emitted from the initial-state electron and positron, can
  combine to produce $e^+e^-$, $\mu^+\mu^-$, or $\pi^+\pi^-$ pairs.
  The electron and positron are usually, but not always, lost along the
  beam pipe.
  A 4-lepton event is obtained if the initial state pair is observed, or if an additional
  $\gamma^*$ produces two more leptons.
  The latter kind of event can be rejected by requiring reconstruction of the full beam energy.
\item Sequential leptonic $B$ decays produce events with up to four
  leptons, plus additional tracks; some energy is carried by neutrinos
  so these events do not reconstruct the full beam energy.
\end{enumerate}

\paragraph{$4\ell$ exclusive, reconstructing $E_{\rm cm}$ (with or without $\gamma$):}
Four-lepton final states ($\mu^+\mu^-$ $\mu^+\mu^-$, $\mu^+\mu^-e^+e^-$,
and $e^+e^-e^+e^-$) that reconstruct the full beam energy can be
produced by the simplest decay chains in Higgsed sectors (Figure \ref{fig:feynHiggsed}(a)), or by
confined sectors with $m_{\eta'_D} \lesssim E_{\rm cm}/4$.  These
processes also produce $\tau$ and hadronic final states, but these are
harder to search for, whereas muons have the lowest backgrounds.
Requiring the reconstruction of the full beam energy $E_{\rm cm}$ removes
the physics backgrounds 2) and 3) mentioned above, but also makes
these searches insensitive to production modes that include stable or
long-lived final states, such as baryons or metastable Higgs bosons.
Pairs of leptons should reconstruct very narrow (resolution-limited)
resonances, which can be used as a very useful discriminator between
signal and background.  The two lepton pairs need not reconstruct the
same mass, but it is likely that they will reconstruct
indistinguishable masses in an $\OO$(1) fraction of events.  Similarly,
$4\ell+\gamma$ final states can be produced by simple Higgsed-sector
decays, or by confined sectors with $m_{\eta'_D} \lesssim m_{A'}/4$
and $m_{A'} < E_{\rm cm}$.  In this case, in addition to pairwise mass
reconstruction as in the pure $4\ell$ case, the four leptons should
now also reconstruct a (potentially broader) $A'$ resonance.

\paragraph{$4 \ell$ exclusive with displaced vertices (with or without $\gamma$):}
The requirement of displaced vertices allows significant reduction of
Standard Model backgrounds 1) and 2), without demanding that the
observed products reconstruct the full beam energy $E_{\rm cm}$.  In
an exclusive four-lepton final state, background 3) is also negligible.
Dropping the $E_{\rm cm}$ requirement is quite important --- in
scenarios with displaced vertices, it is also generic to have missing
energy coming from long-lived particles decaying outside the detector,
or exactly stable baryons.  For example, a confined sector with one
light flavor can have displaced $\omega_D$ decays, and long-lived
invisible $\eta'_D$ decays. Similarly, the beam-energy requirement can
be relaxed in a $4\ell+\gamma$ search and replaced by a displaced
vertex cut.  In this case, a four-lepton reconstructed mass peak is
\emph{no longer} expected as some of the $A'$ mass can go into
invisible decays.  Pairs of same-flavor leptons should still
reconstruct narrow resonances.

\paragraph{$\ge 5 \ell$ exclusive (with $E_{\rm cm}$ or displaced
  vertices, with or without $\gamma$):}
More complex Higgsed-sector decay chains such as Figure \ref{fig:feynHiggsed}(a), where the
Higgs decays within the detector, or confined-sector decays with
larger ratio $E_{\rm cm}/\Lambda_D$ can generate moderate track
multiplicities.  We stress that, though the leptonic final states are
most distinctive, especially at higher masses the dark resonances
\emph{will} decay to $\tau$ and hadronic states as
well.  Therefore, it is essential not to veto on non-leptonic tracks,
and to tag leptons as loosely as possible to suppress backgrounds.  It
may also be interesting to specifically search for events with $\tau^+\tau^-$
or $\pi^+\pi^-$ pairs.   Therefore, a more
inclusive approach is warranted, i.e. requiring 5 or 6 leptons to
reduce background, plus additional tracks.
As in the $4\ell$ searches, two complementary
cuts are useful for discriminating from background: either full
reconstruction of $E_{\rm cm}$ or displaced vertices.

\paragraph{High Track Multiplicities}
A confined dark sector with $m_{\eta'_D} \sim m_{\eta'_{QCD}}$
produces a high multiplicity $\sim 10-20$ dark hadrons, of which a
significant fraction decay visibly within the detector.  The resulting
dark-sector parton shower can lead to somewhat collimated jets of
decay products.  A large fraction (between ~4/5 and 1/3 depending on
meson mass) of dark mesons go into $e$ or $\mu$ pairs, and this can be
used to assist in distinguishing these final states from backgrounds.
There is no source of analogous events from a Higgsed dark sector
without a confined group.

\paragraph{$\gamma$ + nothing:}
A further interesting region of parameter space is when all the
produced dark-sector states decay outside the detector, in which case
all dark-sector products are invisible.  This occurs, for example, if
all states in the dark sector can decay to a long-lived pion or dark
Higgs.  In this case, the `$\gamma$ + nothing' mode is the only effective search at
low-energy $e^+e^-$ colliders (see also \cite{Borodatchenkova:2005ct}).  
These events can be detected by observing a peak in the photon energy 
spectrum at $\sqrt{s}\simeq m_{A'}$.

\vskip 4mm

The searches mentioned above are complementary
to the dilepton resonance searches
($e^+e^- \to \gamma A' \to \gamma \ell^+\ell^-$) that have recently been discussed in 
\cite{Borodatchenkova:2005ct,Pospelov:2008zw,Aubert:2009cp}.
However, the $A' \rightarrow \ell^+\ell^-$ decay is significant only when the
$A'$ is the lightest state in the dark sector.
Even if the $A'$ is the
lightest state of the dark sector, any other kinematically accessible
states are pair-produced with a comparable rate through an off-shell
$A'$, which produces a signal in the multi-lepton
channel, which has much lower QED backgrounds.  The multi-lepton
searches considered here overlap with the 4- and 6-lepton searches
discussed in \cite{Batell:2009yf} in the context of Abelian dark
sectors, but we emphasize that much higher multiplicities are easily
obtained in more general non-Abelian dark sectors.

\section{Conclusions}\label{sec:conclusion}
Evidence for a low-mass dark sector may be hidden in unexamined $\epm$
collider data.  
If the new sector contains a
$U(1)_D$ factor, then moderately large kinetic mixing $\epsilon \sim
10^{-3}$ is not only generated by generic mechanisms, but suggested by
a variety of experiments and observations that look for dark matter.
States of a new sector with couplings in this
range can be efficiently produced in $e^+e^-$ colliders, and may decay
spectacularly, so that potential discoveries are within
the reach of searches in existing data from low-energy $e^+e^-$
collider experiments such as BaBar, BELLE, CLEO-c, and KLOE.

The evidence for a ``dark sector'' from terrestrial and satellite
searches for dark matter is indirect, but strikingly consistent.  The
large local electron/positron excesses reported by HEAT, PAMELA,
PPB-BETS, and ATIC suggest interactions of $\OO$(TeV)-mass dark matter
with a new boson with mass $\OO$(1 GeV).  Simultaneously, two
experiments suggest multiple dark matter states with small mass
splittings: the INTEGRAL 511 keV-line and DAMA/LIBRA modulation signal
are consistent with $1-10$ MeV and $\sim 100$ keV splittings among
dark matter states, respectively.  New forces, Higgsed or confined at
$\OO$(GeV), can give rise to these splittings.  Any subset of these
results suggests a departure from the standard minimal WIMP paradigm,
but dark matter data will provide at best an indirect probe of such
structure.

If the DAMA/LIBRA modulation signal comes from inelastic scattering
mediated by kinetic mixing, then it is the first quantitative probe of
kinetic mixing between the Standard Model and the dark sector, and can
be used to estimate the production rate of new dark-sector particles.
If the mass of the new $U(1)_D$ gauge boson is $m_{A'} \simeq$ 1 GeV,
cross-sections $\simeq$ 0.1 fb are expected in a Higgsed dark sector
with $\alpha_D \simeq \alpha$.  This estimate leads to 50-100 events
expected in the stored datasets from BaBar and BELLE.  The $\epm$
production cross-section normalized to DAMA/LIBRA scales as
$m_{A'}^4$, so much larger cross-sections ($\OO$(pb)) are possible for
a heavier $A'$, while higher-luminosity experiments are required to
observe dark sectors with lower $m_{A'}\sim 100$'s of MeV.
Cross-sections up to 100 fb are possible even for a 1 GeV $A'$, if the
DAMA/LIBRA scattering is suppressed by a form factor, as expected in
confined models.

Though spectacular, the signatures of dark-sector production at
$e^+e^-$ colliders would still have been missed by existing searches.
The precise final states depend on the spectroscopy of the dark sector
--- general regions of parameter space allow a combination of prompt
and displaced decays, 
and states that escape the detector
invisibly before decaying.  All decay modes are expected to produce
lepton-rich final states.  Moderate multiplicities (4-10) of leptons
and very high multiplicities are both possible, and require different
search strategies.  Some of these lepton pairs are expected to
reconstruct very narrow resonances, which in principle can be combined
to reconstruct the spectrum of the dark sector.  Dark sectors with
very long-lived final states may only be visible in a
`$\gamma+\mbox{nothing}$' search.  Further work to optimize the
strategies proposed here and study their efficiencies is an important
task.

The small production cross-sections for $m_{A'}\lesssim 100$ MeV pose
a particular challenge.  Novel experiments are called for to
explore this region.  Two features of the low-mass region are
noteworthy.  Lower-energy $\OO(100 \mbox{ MeV})$ colliders may reach
higher luminosities than can be attained at GeV-energy colliders.
Moreover, not far below the cross-section reach of B-factories, at
$\epsilon \sim 10^{-6}$, the expected decay lengths of spin-1 states
in the dark sector become $\OO$(1 m), so that very different
techniques, such as beam-dump experiments, may be the most fruitful
way to isolate dark-sector production events.

Dark matter experiments suggest new low-energy gauge interactions
beyond the Standard Model.  If a dark sector exists, it dramatically
reshapes our understanding of the structure of nature.  Searches in
existing collider data offer a simple but powerful probe of this new
dynamics and should be pursued.

\subsubsection*{Acknowledgments} We are grateful to
Daniele Alves, Nima Arkani-Hamed, Siavosh Behbahani, Howard Haber, Jay
Wacker, and Neal Weiner for many fruitful discussions. We especially
thank Mathew Graham and Aaron Roodman for numerous discussions
regarding BaBar physics, and Michael Peskin for helpful feedback on
our analysis of production and decay phenomenology. RE and PCS are
supported by the US DOE under contract number DE-AC02-76SF00515.

\appendix

\section{Constraints on Kinetic-mixing between hypercharge and a new $U(1)_D$}\label{app:Contraints}

In this appendix, we review the most important existing constraints on the kinetic mixing
between Standard Model hypercharge and $U(1)_D$.

For low-mass (MeV--GeV) dark photons $A'$, there are a variety of constraints on the kinetic mixing that are
summarized in \cite{Pospelov:2008zw} (for a discussion of searches for even lower mass dark photons see 
\cite{Ahlers:2007qf,Ahlers:2008qc}).
Constraints from Big-Bang Nucleosynthesis disfavor $m_{A'}\lesssim\OO$(few MeV) \cite{Serpico:2004nm}.
For $m_{A'} \lesssim 10$ MeV, the strongest constraints come from the $A'$ contribution to the anomalous
magnetic moment of the electron, $a^{A'}_e$, while for $m_{A'}$ above 10 MeV, but less than a few GeV, the tightest constraints
come from the $A'$-contribution to the anomalous magnetic moment of the muon, $a^{A'}_\mu$.
For $m_{A'} \sim \OO$(100 MeV - 300 MeV), constraints from certain rare decays of mesons,
such as $K^+\to \pi^+ A'$, are in some cases competitive with the constraint on
$a^{A'}_{\mu}$, but depend sensitively on the decay modes of the $A'$ --- see \cite{Pospelov:2008zw}
for a more detailed discussion.
Above a few GeV, kinetic mixing is only very weakly constrained.
However, since kinetic mixing also induces a coupling between the $Z$ and
dark-sector matter, a potential probe could come from
searches for rare $Z$-decays.

Since we are interested in the mass range $m_{A'} \sim $ 10 MeV - 20 GeV, we focus on
the constraint on $a^{A'}_\mu$ and the potential probe coming from rare $Z$-decays.

\paragraph{Anomalous Lepton Magnetic Moments:}
Let us first discuss the contribution from the $A'$ to the anomalous magnetic moment of
a lepton, $\ell$.
This was calculated in \cite{Pospelov:2008zw}, and is given by
\be\label{eq:muon}
a^{A'}_\ell = \f{\alpha \epsilon^2}{2\pi} \, \int_0^1dz \f{2m^2_\ell z(1-z)^2}{m_\ell^2 (1-z)^2 + m_{A'}^2 z}.
\ee
For $a^{A'}_e$, this can be converted to a constraint of $\epsilon^2 \lesssim 10^{-5} (m_{A'}/(10 \;{\rm MeV}))^2$
\cite{Pospelov:2008zw}, which is important for $m_{A'}\lesssim 10$ MeV.
A constraint on $a^{A'}_\mu$ is more ambiguous, since the theoretical prediction for the muon anomalous magnetic
moment within the Standard Model, $a_\mu$, involves a hadronic contribution which must be estimated from
experiments, which do not all agree.
Using data from $e^+e^-$ annihilation to hadrons, the theoretical value of $a_\mu$ is smaller than the
measured value by $(302\pm 88)\times 10^{-11}$, a $3.4\sigma$ deficit \cite{Passera:2008hj,Bennett:2006fi}.
However, estimates involving data from $\tau$'s \cite{Davier:2002dy} or from preliminary BaBar results on the
precision measurement of the $e^+e^-\to\pi^+\pi^-(\gamma)$ cross-section with the ISR method
\cite{Davier08} do not point to a significant discrepancy.
Since the $A'$ contribution to $a_{\mu}$ is positive, a conservative constraints on the $\epsilon-m_{A'}$ parameter space
is obtained by taking the estimate from $e^+e^-$ annihilation to hadrons.
In particular, for the constraints shown in Section \ref{sec:production},
we follow \cite{Pospelov:2008zw} and require $a_\mu^{A'} < (302+5\sigma)\times 10^{-11} = 7.4\times 10^{-9}$.
For $m_{A'}\gg m_\mu$, equation (\ref{eq:muon}) reduces to $a^{A'}_\mu \simeq \f{\alpha \epsilon^2 m_{\mu}^2}{3\pi m_{A'}^2}$,
and the constraint on $a^{A'}_{\mu}$ thus becomes very weak for
$m_{A'}$ larger than a few GeV.

\paragraph{Exotic $Z$-Decays:}
For $m_{A'}$ larger than $m_\mu$, searches for exotic $Z$-decays among the $2\times 10^7$ $Z$-bosons recorded by the experiments at LEPI
could potentially probe kinetic mixings more sensitively than $a_{\mu}$.
Let us discuss this in more detail.
The coupling induced between the $Z$ and dark-sector fermions is given by
\be
\mathcal{L} \supset - \sum_{i=1}^{N_f} \epsilon g_D \tan\theta_W Z_{\mu} \bar{X} \gamma^{\mu} X,
\ee
where $N_f$ is the number of dark-sector particles coupling to $A'$ with gauge coupling $g_D$, and
$\sin\theta_W$ is the Standard Model weak-mixing angle.
In this expression, the $A'$ coupling to the electromagnetic current has been normalized as
$\mathcal{L}\supset \epsilon g_D A'_{\mu} J^{\mu}_{\rm EM}$, see equation (\ref{eq:mixing}).
The $Z$-decay width to dark-sector fermions is thus
\be
\Gamma(Z \to \bar{X} X) \sim \f{1}{3} N_c \epsilon^2 \alpha_D m_Z \tan^2\theta_W\sum_{i=1}^{N_f}
 \sqrt{1-\f{4 m_X^2}{m_Z^2}} \Big(1+\f{2 m^2_{X}}{m_Z^2}\Big),
\ee
where $N_c$ is the number of colors of the dark-sector gauge group.

Since the total width of the $Z$ has been measured to be $2.4952 \pm 0.0023$ GeV \cite{Amsler:2008zzb}, the branching ratio
of $Z\to \bar{X} X$ is given by
\be\label{eq:Zdecay}
{\rm BR}(Z\to\bar{X}X) \sim 2.8\times 10^{-8} \, N_c \, \f{\alpha_D}{\alpha}\, \Big(\f{\epsilon}{10^{-3}}\Big)^2,
\ee
where we assumed $m_X \ll m_Z$.
Depending on the decay modes of $X$, which we discussed in detail in Section
\ref{sec:generalSector}, a variety of signatures are possible.
For example, if the $X$'s decay invisibly or
outside the detector, the constraint is not better than the
uncertainty on the width of the $Z$, which is about $0.1\%$.  More
exotic $Z$ decays into multi-leptons are possible, but require
dedicated searches.  The best constraints on rare $Z$ decay branching
ratios are no better than the $10^{-6}-10^{-5}$ level \cite{Amsler:2008zzb}, and
even dedicated searches may not reach this level for complicated final
states.
Equation (\ref{eq:Zdecay}) shows that for typical parameters that
give $\OO$(fb) cross-sections at B-factories, the branching ratio
$Z\to\bar{X}X$ is about three orders of magnitude smaller than what could
be probed at LEPI even with a dedicated search.
For the multi-lepton final states we consider in this paper, the B-factories
will thus be a far more sensitive probe of the $\epsilon-m_{A'}$ parameter space than
LEPI, at least for $m_{A'}$ less than a few 10's of GeV.

\section{DAMA/LIBRA Normalization of Couplings}\label{app:iDMreview}

Inelastic dark matter offers a simple explanation that reconciles the annual modulation
signal reported by DAMA/LIBRA with the null results of other experiments, and is reviewed in \cite{TuckerSmith:2001hy,Chang:2008gd}.
Below, we summarize our procedure for normalizing $\alpha_D \epsilon^2$
assuming that DAMA/LIBRA is explained by an iDM mechanism.
We will normalize $\alpha_D \epsilon^2$ for two iDM scenarios:
one in which the inelastic channel proceeds through charged-current scattering as in \cite{TuckerSmith:2001hy},
and one where the inelastic channel is through a hyperfine transition as in \cite{Schuster:2009}.

For charged-current scenarios, the inelastic cross-section to scatter off of a
nucleus of mass $m_N$ and charge $Z$ recoiling with energy $E_R$ is
\bea
\frac{d\sigma}{dE_R} &\approx& \frac{8\pi Z^2\alpha \alpha_D \epsilon^2 m_N}{v^2} \frac{1}{(2m_NE_R+m_{A'}^2)^2} |F(E_R)|^2, \label{eq:ChargedScattering}
\eea
where $\alpha$ is the fine structure constant, $v$ is the dark matter nucleus relative velocity, and $m_{A'}$ is the $A'$ mass.  $F(E_R)$ is the Helm nuclear form factor \cite{Helm:1956zz}, and it accounts for the loss of coherence scattering off of the entire nucleus at large recoil. We use,
\be
|F(E_R)|^2 = \left(\frac{3j_1(|q|r_0)}{|q|r_0}\right)^2 e^{-s^2|q|^2},
\ee
where $s=1$ fm, $r_0=\sqrt{r^2-5s^2}$, $r=1.2A^{1/3}$ fm, and $|q|=\sqrt{2m_NE_R}$.

For hyperfine transition scattering \cite{Schuster:2009}, the inelastic cross-section is
\bea
\frac{d\sigma}{dE_R} &\approx& \frac{4\pi Z^2\alpha c_{in}^2\alpha_D \epsilon^2}{v^2\Lambda_D^2}\frac{m_N^2 E_R}{(2 m_N E_R+m_{A'}^2)^2}|F(E_R)|^2, \label{eq:DipoleScattering}
\eea
where $c_{in}/\Lambda_D \sim \text{GeV}^{-1}$ is the $A'$ electric dipole of the dark matter.
To fix the relation between $\Lambda_D$ and the hyperfine splitting $\delta$ in this case,
we introduce a scale $\tilde{\Lambda}_D$ such that
\bea
\delta &=& \frac{\tilde{\Lambda}_D^2}{m_{DM}}.
\eea
As discussed in Section \ref{sec:DAMANormalization}, $\tilde{\Lambda}_D/\Lambda_D$ is naturally a small number.
We will therefore normalize $\alpha_D\epsilon^2$ as a function of $m_{A'}$ for different choices of the small dimensionless coefficient,
\bea
\tilde{C}=\left(\frac{c_{in}^2\tilde{\Lambda}_D^2}{\Lambda_D^2}\right).
\eea

Following \cite{Lewin:1995rx}, the differential scattering rate per unit detector mass is,
\bea
\frac{dR}{dE_R} = \frac{\rho_0 v_0}{m_N m_{DM}} \int_{v_{\text{min}}(E_R)}\hspace{-0.3in}d^3 v\hspace{0.1in} \frac{v}{v_0} f(v;v_e) \frac{d\sigma}{dE_R},
\eea
where $\rho=0.3$ GeV/cm$^3$ is the local density of dark matter, and $f(v)$ is the dark matter velocity and velocity distribution function in the lab frame. Introducing a $\Theta(v_{\text{esc}}-v)$ to naively cut off the velocity profile above the galactic escape velocity $v_{\text{esc}}\approx 500-600$ km/s, we use a simple velocity distribution function,
\bea
f(v;v_e) \propto \Big(e^{ - \frac{(\vec{v} - \vec{v}_e)^2}{ v_0^2}}\! - e^{- \frac{v_{\text{esc}}^2}{v_0^2}} \Big)
 \Theta(v_{\text{esc}}-|\vec{v}-\vec{v}_e|),
\eea
where $v_0\approx 220$ km/s, and $v_e$ is the earth's speed in the galactic frame. We use the numerical values reported in \cite{Lewin:1995rx} for $\vec{v}_e$.
For the annual modulation rate, we use
\bea
\bigg(\frac{dR}{dE_R}\bigg)_{\text{modulated}}\approx \frac{1}{2}\bigg(\frac{dR}{dE_R}\Big|_{\text{June 2}}-\frac{dR}{dE_R}\Big|_{\text{Dec 2}}\bigg).
\eea
To normalize the combination $\alpha_D\epsilon^2$, we fit to the reported DAMA/LIBRA signal, using the combined 0.82 ton-yr exposure\cite{Bernabei:2005hj}.
We do this by minimizing a $\chi^2$ function using the 12 half-keVee bins and reported error bars between $2-8$ keVee with $10=12-2$ independent degrees of freedom.
The lower and upper ranges of $\alpha_D\epsilon^2$ are taken by
scanning over the region with $\chi^2\le 16$, and typically differ by
a factor of $\sim 10$.
Our estimate is obtained by taking the geometric mean of the lower and upper values of $\alpha_D\epsilon^2$.

\bibliographystyle{JHEP}   
\bibliography{darkSectorPaper}
\end{document}